\documentclass[12pt]{article}

\usepackage[utf8]{inputenc} 
\usepackage{hyperref}       
\usepackage{url}            
\usepackage{booktabs}       
\usepackage{amsfonts}       
\usepackage{nicefrac}       
\usepackage{microtype}      
\usepackage{lipsum}
\usepackage{graphicx}
\usepackage{moreverb}
\usepackage{enumitem}
\usepackage{upgreek}
\usepackage{subcaption}
\usepackage{mathtools}
\usepackage{adjustbox}
\usepackage{tabularx}
\usepackage{bbm}
\usepackage{colortbl}
\usepackage{listings}
\usepackage[superscript,biblabel]{cite}

\newcommand{\pa}[2]{\alpha_{#1}^{\MakeUppercase{#2}}}
\newcommand{\pb}[2]{\beta_{#1}^{\MakeUppercase{#2}}}
\newcommand{\pk}[1]{k_{t}^{\MakeUppercase{#1}}}

\newcommand{\fl}[1]{\gamma_{#1}^{L}}
\newcommand{\fr}[1]{\gamma_{#1}^{R}}
\newcommand{\fk}[1]{k_{f}^{\MakeUppercase{#1}}}
\newcommand{\fs}[2]{\sigma^{\MakeUppercase{#1}}_{#2}}

\usepackage[a4paper,left=0.75in,right=0.75in,top=1in,bottom=1in]{geometry}
\PassOptionsToPackage{normalem}{ulem}
\usepackage{ulem}
\usepackage{xcolor}
\usepackage{amsthm}
\theoremstyle{definition}

\newtheorem{assumption}{Assumption}
\newtheorem{theorem}{Theorem}
\begin{document}
\sloppy

\def\spacingset#1{\renewcommand{\baselinestretch}%
{#1}\small\normalsize} \spacingset{1}

\title{Bayesian regression discontinuity design\\ with unknown cutoff}

 \author{Julia Kowalska\footnote{Corresponding author, E-mail address: j.m.kowalska@vu.nl} \textsuperscript{1,2,3}, Mark van de Wiel\textsuperscript{1,2}, Stéphanie van der Pas\textsuperscript{1,2,3}\\
\textsuperscript{1}\small{Amsterdam UMC location VU, Epidemiology and Data Science, Amsterdam, The Netherlands}\\
\textsuperscript{2}\small{Amsterdam Public Health, Methodology, Amsterdam, The Netherlands}\\
\textsuperscript{3}\small{VU Amsterdam, Department of Mathematics, Amsterdam, The Netherlands}\\
}
\maketitle

\begin{abstract}
The regression discontinuity design (RDD) is a quasi-experimental approach used to estimate the causal effects of an intervention assigned based on a cutoff criterion. RDD exploits the idea that close to the cutoff units below and above are similar; hence, they can be meaningfully compared. Consequently, the causal effect can be estimated only locally at the cutoff point. This makes the cutoff point an essential element of RDD. However, the exact cutoff location may not always be disclosed to the researchers, and even when it is, the actual location may deviate from the official one. As we illustrate on the application of RDD to the HIV treatment eligibility data, estimating the causal effect at an incorrect cutoff point leads to meaningless results. The method we present, LoTTA (Local Trimmed Taylor Approximation), can be applied both as an estimation and validation tool in RDD. We use a Bayesian approach to incorporate prior knowledge and uncertainty about the cutoff location in the causal effect estimation. At the same time, LoTTA  is fitted globally to the whole data, whereas RDD is a local, boundary point estimation problem. In this work we address a natural question that arises: how to make Bayesian inference more local to render a meaningful and powerful estimate of the treatment effect? 
\end{abstract}
\noindent%
\vfill

\spacingset{1}
\section{Introduction}\label{sec1}

The regression discontinuity design (RDD) is a quasi-experimental approach used to estimate the causal effects of an intervention assigned based on a cutoff criterion. It was applied for the first time in 1960 by Thistlethwaite \& Campbell to evaluate the effect of receiving a certificate of merit on students' future academic careers. \cite{Thistlethwaite1960} Despite its increasing popularity in empirical studies, and many opportunities, RDD appears to be a common design in econometrics, but it is still underutilized in other disciplines such as medicine. \cite{Moscoe2015}$^,$\cite{Hilton2021} 

RDD corresponds to the following framework.\cite{Imbens2008}$^,$ \cite{Cattaneo2023} First, each unit receives a \emph{score}.  The score is a pre-intervention variable, such as age or blood pressure, which often influences the intervention's outcome. Next, the score is compared to the prespecified \emph{cutoff value}.  The cutoff criterion may be the only determinant of the intervention allocation, in which case the design is called \emph{sharp}. In the sharp design all units with scores equal or higher than the cutoff value are assigned to one type of an intervention, and all units with scores below to the other. If the compliance to the cutoff rule is imperfect, then the design is called \emph{fuzzy}. In this scenario units with scores just above the cutoff have significantly higher chances of being assigned to the intervention than units with scores just below. Finally, for each unit an outcome of interest is observed. RDD exploits the idea that (infinitesimally) close to the cutoff, units below and above are similar in terms of important characteristics; hence, they can be meaningfully compared. \cite{Lee2008} In particular, one of the strengths of RDD is that it does not require covariate adjustment to identify causal effects.\cite{Cattaneo2023covariates}  By estimating the treatment effect at the cutoff, the common problem of unbalanced groups is mostly avoided at the price of restricted generalizability. 

We speculate that the underutilization of RDD outside econometrics   is partially explained by the imperfect compliance to many  interventions, necessitating the less well-studied fuzzy design. A further complication appears when the cutoff cannot be disclosed to the researchers. This may occur when a cutoff is part of a protected internal policy - such as those used by schools \cite{Bütikofer2023,Brunner2023,Landaud2020,Cotofan2022} or banks\cite{Fukui2023,Agarwal2018} - or if the precise cutoff values were never recorded or are no longer retrievable.  For example, this issue may arise in  registries that collect data from many anonymized hospitals,\cite{Steenbergen2015} in records of historical admission thresholds across educational institutions\cite{Mountjoy2024,Hoekstra2009,Booij2016} or in documentation of policies' implementation dates. \cite{Zaniewski2021} In many of those cases, researchers may have partial knowledge about the cutoff location, making the setting well-suited to a Bayesian approach. 

In addition, attention should be drawn to a similar problem of \emph{seemingly} known cutoffs. A commonly held assumption is that the cutoff given by an official guideline is the same point at which the treatment effect is to be estimated. However, those are not necessarily the same, as will become apparent in our reanalysis of ART data from Hlabisa HIV Treatment and Care Programme. 

Currently, the standard method of analyzing an RDD with a known cutoff is through weighted local linear regression (LLR). It is a well-studied framework that provides robust,  asymptotically valid confidence intervals. \cite{Calonico2014} Moreover, some methods were developed to estimate the location of an unknown cutoff. \cite{Porter2015} However, once the cutoff point is estimated, it is treated as known; we are not aware of any available method that allows us to include the uncertainty about its location in the final treatment effect estimation. In this paper, we present a method that aims to be both an estimation and a falsification tool for RDD's with unknown or uncertain cutoffs. By turning to the Bayesian paradigm, the uncertainty about the cutoff is included in a natural way, as part of a larger model. Furthermore, expert knowledge about the suspected cutoff location can be easily incorporated through an informative prior. 

Bayesian methods have previously been designed for RDD's with a known cutoff. Geneletti et al.\cite{Geneletti2015} fit Bayesian linear regression in a cutoff's neighborhood, which ultimately ought to be selected through expert-knowledge. The main focus is placed on the prior selection and sensitivity analysis. Chib et al. \cite{Chib2023} also emphasize the influence of points near the cutoff, but do not discard points further away. Instead, they fit cubic splines globally with a higher density of knots in the cutoff's neighborhood selected based on the score's quantiles. Finally, Branson et al.  \cite{Branson2019} propose the use of a flexible nonparametric model with an exponential Gaussian process prior. Unlike the aforementioned  solutions, no emphasis is put on any particular neighborhood of the cutoff.  The inference is stabilized by shared covariance parameters that are estimated from the data on both sides of the cutoff. The above methods cannot be easily translated to the unknown cutoff scenario, as they require manual adjustments that depend on the cutoff location or as they are challenging to implement in an efficient way.

Thus, our main contribution is the development of a fully Bayesian approach to RDD with an unknown cutoff. To the best of our knowledge our model, LoTTA (Local Trimmed Taylor Approximation), is the first one in the RDD literature that propagates the uncertainty of the cutoff location to the treatment effect estimation. A significant advantage of the model is that it does not require tuning or manual window selection, making it easy to apply. We also contribute a new perspective on feedback between design and analysis stages in Bayesian causal inference,\cite{Zigler2013}$^,$\cite{Robins2015}$^,$\cite{McCandle2010} as in contrast to propensity score based models, we find significant advantages towards allowing such feedback in RDD.  Even in settings where the cutoff is known, LoTTA turns out to be a useful validation tool for the cutoff location, as demonstrated in Section \ref{diagnostic}. We demonstrate that the current methods may be insufficient and explain how the Bayesian model can help to fill this gap.

The paper is organized as follows. In Section \ref{setup}, we introduce the notation and present the causal inference framework for RDD. In Section \ref{model}, we discuss the cutoff detection model in detail, along with comments on its implementation and its place in the wider context of causal inference. In Section \ref{diagnostic}, we explain through a real-life example the steps of the cutoff detection analysis and how the model can be used as a falsification tool. We conclude this section with a surprising finding of a cutoff located at a different point than implied by the corresponding guidelines. In Section \ref{simulations}, we present simulation results on the frequentist properties of the model. Finally, in Section \ref{discussion}, we summarize key takeaways.  
\section{Setup}\label{setup}
\subsection{Unknown cutoff}
Throughout the paper we assume that decision makers, while assigning units to the interventions, were guided by one fixed cutoff that was known to them. While commonly the researchers analyzing the
data know the cutoff value used by the decision makers,
this is not necessarily the case. A motivating example comes from the Dutch Arthroplasty Register\cite{Steenbergen2015}, which collects data on joint replacements from all Dutch hospitals. Some Dutch hospitals use an age-based cutoff criterion to decide on the fixation type, and each hospital is free to select a cutoff value, if they use one at all. A research team wishing to exploit the age-based cutoff for an RDD analysis was hindered by the inability of the Register to supply them with identifying information for the hospitals at which patients were treated.\cite{Graeff2024} The researchers were therefore unable to find out which cutoff values were used for which patients in their data set.

The method we present, although designed for a fixed but unknown cutoff, is also a useful validation tool for a \emph{seemingly} known cutoff. The current methods focus on the validation of the design at one point, potentially missing or misinterpreting evidence of a cutoff being set to a different value than given by the guidelines. As we show on a real data example in Section \ref{diagnostic}, this may lead to incorrect validation of an RD design.
\subsection{Causal framework}
In terms of causal inference we reason in the potential outcomes framework with two competing treatments. \cite{Hernan2020} This framework assumes that for each unit there exist two outcomes corresponding to each treatment but we can only observe one of them - the one of the assigned treatment. Throughout the paper, we denote the potential outcomes under treatment and no treatment by $Y^{(1)}$ and $Y^{(0)}$ respectively. We focus on fuzzy RDD, so on the scenario of imperfect compliance to the cutoff rule. In a sharp design the cutoff is straightforward to detect. Moreover, we work within the continuity framework as opposed to the local randomization framework.\cite{Cattaneo2015,Li2015} Although the local randomization framework offers improved generalizability of results, it does not readily apply to scenarios with an unknown cutoff. In particular, existing methodologies rely on balance tests across multiple covariates, which are not straightforward to implement without precise knowledge of the cutoff location.  

As is usual in fuzzy RD,\cite{Imbens2008} we differentiate four subgroups of participants based on their compliance type ($C_T$): alwaystakers (A), nevertakers (N), compliers (C) and defiers (D). Alwaystakers and nevertakers, respectively, either always or never receive the treatment regardless of their score. Compliers receive the treatment according to the cutoff rule based solely on their score, while defiers receive the opposite treatment with respect to the cutoff rule. The causal estimand of interest is the local  average treatment effect $\tau$, which in this case is the expected mean difference between $Y^{(1)}$ and $Y^{(0)}$ for compliers with scores equal to the cutoff value, see \eqref{tau}.

\subsection{Notation}\label{notation}
In this section we introduce the notation and some general assumptions. By $X$ we denote the score variable that takes values in a bounded interval $\mathcal{I}=[I_1,I_2]$; by $Y$ the outcome variable that takes values in $\mathbb{R}$; by $T\in\{0,1\}$ the binary treatment received; and by $c\in (I_1,I_2)$ the cutoff location. To identify the treatment effect $\tau$, we make a standard assumption of no defiers .\cite{Imbens2008}
  \begin{assumption}\label{as1}
      For each unit $\mathbb{P}(T_i|X=x_1)\leq \mathbb{P}(T_i|X=x_2)$, given $x_1<x_2$.
  \end{assumption}
Furthermore, we consider two functions: treatment probability function $p$ and outcome function $f$. 
\begin{assumption}\label{as2}
   The treatment probability function $p\colon \mathcal{I}\rightarrow [0,1]$ has precisely one discontinuity point at $c$, and is given by the relation
\begin{align*}
    \mathbb{P}(T=1|X=x)=p(x).
\end{align*}
Moreover, at $c$ the function $p$ is right-continuous, and the conditional probability functions $\mathbb{P}(C_T=c_t|X=c) $ are continuous in $x$ for each compliance type $c_T=A,N,C$.  
\end{assumption}
 \begin{assumption}\label{as3}
     There exists a right-continuous deterministic function $f$ with at most one discontinuity point at $c$ such that
\begin{align*}
    Y|_{X=x}\sim \mathcal{N}(f(x),\sigma_x).
\end{align*}
Moreover, the conditional expected values $\mathbb{E}[Y^{(i)}|X=x,C_T]$  are continuous in $x$ on the whole domain for each potential outcome $i=0,1$ and compliance type $C_T=\text{A}, \text{N}, \text{C}$.
 \end{assumption}
In the model we consider in Section \ref{model}, we allow the error term to depend on the score $x$, however not in full generality. 

Assumptions \ref{as1}, \ref{as2}, and \ref{as3} ensure identification of the local average treatment effect:\cite{Imbens2008} 
\begin{align}
    \begin{split}\label{tau}
        \tau\coloneqq \mathbb{E}[Y^{(1)}-Y^{(0)}|X=c,\text{C}]=\frac{\lim_{x\downarrow c}\mathbb{E}[Y|X=x]-\lim_{x\uparrow c}\mathbb{E}[Y|X=x]}{\lim_{x\downarrow c}\mathbb{P}[T=1|X=x]-\lim_{x\uparrow c}\mathbb{P}[T=1|X=x]}\\
    =\frac{f(c)-\lim_{x\uparrow c}f(x)}{p(c)-\lim_{x\uparrow c}p(x)}.
    \end{split}   
\end{align}
Assumption \ref{as3} is stronger than strictly necessary for the identification result to hold; the Gaussian error is only required for implementing the estimation procedure, and  can be relaxed to accommodate bounded and binary outcomes (see Appendix \ref{App1}).

The denominator in (\ref{tau}) is equal to the compliance rate at the cutoff. It is important to note that although from a theoretical point of view the above formula is valid for any size of the discontinuity in the treatment probability function, in practice, a small compliance rate, arbitrarily close to zero, presents at least two problems. First, if this rate is small, it calls into question the entire design and the meaning of the treatment effect. Second, it makes the estimation of $\tau$ unstable, leading to uninformative results. From now on, we assume the following.
\begin{assumption}\label{as4}
The compliance rate at $c$, given by $$j\coloneqq \left|\lim_{x\downarrow c}\mathbb{P}[T=1|X=x]-\lim_{x\uparrow c}\mathbb{P}[T=1|X=x]\right|,$$ is not lower than $\eta\in(0,1)$. 
\end{assumption}
 The choice of $\eta$ will depend on a the specific setting. For accurate effect estimation, $\eta$ should be set lower than the expected jump size. For stability, it should not be set too close to zero. For medical applications, in the absence of further knowledge, we consider $\eta=0.2$ a reasonable default value.  For example, in the UK HPV vaccination programme the compliance rate was estimated at 55\%;\cite{Ward2024} in the Wales herpes zoster vaccination programme, it was estimated at 47\%;\cite{Eyting2025} and a study on the administration of antenatal corticosteroids found a change in the treatment take-up of approximately 47\%.\cite{Hutcheon2020} We recommend conducting a sensitivity analysis with respect to parameter $\eta$ when estimating the treatment effect. Further guidance is provided in Section \ref{diagnostic}.

 Finally, we would like to point out that even though the formula in Equation \eqref{tau} resembles Wald's estimator, RDD should not be confused with the instrumental variable design. These two designs share some ideological similarities but the assumptions needed to derive the corresponding estimators are vastly different.\cite{Hahn2001}

\section{Bayesian approach to RDD with unknown cutoff}\label{model}
 
The main challenge of combining cutoff detection with treatment effect estimation lies in the global nature of the former and the local nature of the latter. To detect a jump in the treatment probability function, a broader understanding of the relationship between score and treatment take-up is necessary. Conversely, for treatment effect estimation, the focus shifts to a narrow interval around the cutoff. The data closest to the cutoff is the most valuable; intuitively, the further the score is from the cutoff, the less relevant the corresponding outcome becomes. Our model offers a global fit to data, but it stabilizes the inference near potential cutoffs for higher point estimation precision. 
In the next subsections we present the treatment and outcome models separately and then discuss how they are integrated, and how the outcome model may help in detecting the cutoff.  
\subsection{Treatment probability}\label{treatment_subsection}
The  treatment probability function is the key element for both localizing the cutoff and estimating the compliance rate at the cutoff $j$, which is essential for the design validation and treatment effect estimation (see Equation \eqref{tau}). Even though those tasks are closely related, performing both within a single model requires careful considerations.

We first explain why we do not use piecewise constant priors, despite the effectivity of piecewise constant functions in changepoint detection.\cite{Gijbels2004, Barry1993, Yao1984} The reason is that we not only need to detect the cutoff $c$, but also require an accurate estimate of the compliance rate $j$ at the cutoff. 
In RDD it is generally unrealistic to assume that the probability of receiving treatment is constant on both sides of the cutoff and fitting two constant functions to treatment allocation data can lead to highly biased estimates of $j$  and consequently, even more biased estimates of the treatment effect. Nor would a plug-in procedure based on piecewise constant functions to detect $c$ be advisable for this problem; we give a detailed explanation in Section \ref{joinmodel} and Appendix \ref{Plugin}.

On the other hand,  highly flexible models present two issues as well.  First, such models are not suitable for the cutoff detection as the jump in the treatment probability can be well approximated by a steep but continuous increase in the fitted function, making it difficult to identify the discontinuity. Second, because treatment take-up is a binary variable, fitting highly flexible models without appropriate restrictions can lead to volatile estimates. 

We now impose some mild restrictions on the class of potential treatment probability functions, and then define a model that provides a good approximation for functions within this class and also effectively localizes the cutoff point. Since the cutoff on the score guides treatment assignment, we expect that the further to the right of $c$, the more likely an individual is to receive the treatment, and vice versa. Therefore, we assume the following.
\begin{assumption}\label{as5}
The treatment probability function $p$ is monotone.
\end{assumption}
 For simplicity, but without loss of generality, we assume that $p$ is increasing. Additionally, we expect that the convexity does not change on the each side of the cutoff. Indeed, a change in the convexity can imitate a jump in the treatment probability undermining cutoff detection, and even suggest a flawed study. If we do not know the cutoff, we should be more demanding regarding the treatment allocation mechanism as validation of such design is more difficult.
 \begin{assumption}\label{as6}
    The derivatives of restricted treatment probability functions $p|_{x<c}$, $p|_{x\geq c}$ exist and have no inflection points.
\end{assumption}
 
\begin{figure}
    \centering
    \includegraphics[width=\textwidth]{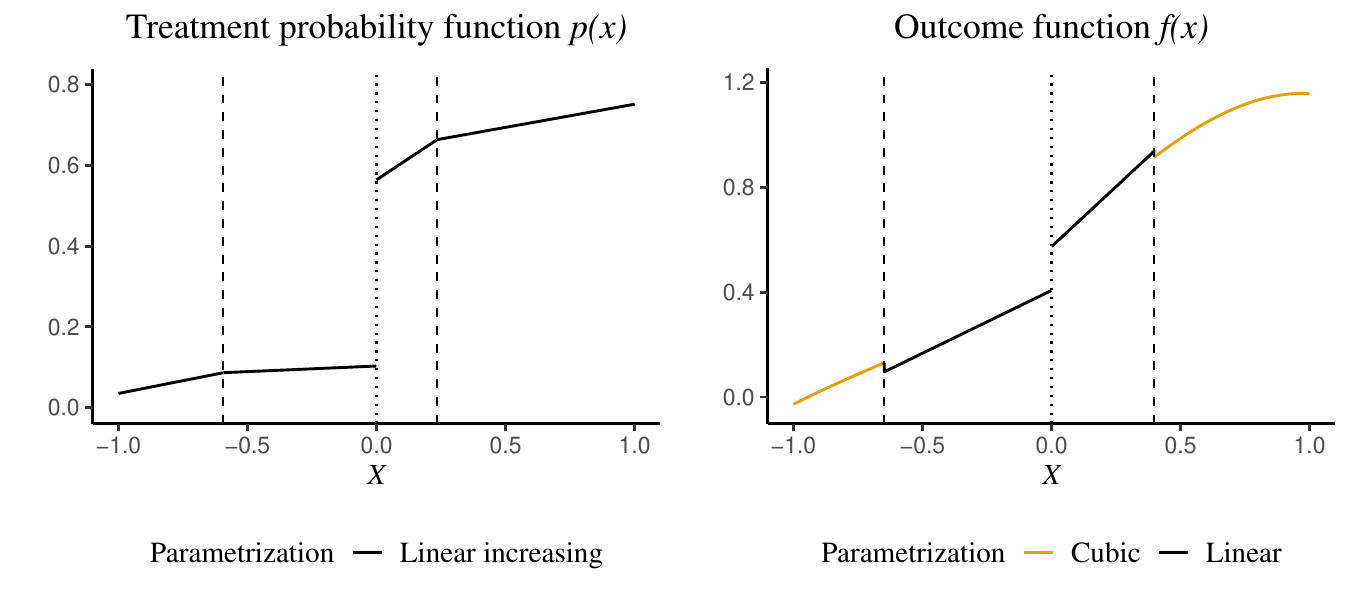}
    \caption{Illustration of the treatment probability function \eqref{pmodel} and the outcome function \eqref{fun_out} with the cutoff at 0.}
    \label{fig:model}
    
\end{figure}

 Having in mind the above considerations, we model the function $p$ on each side of the cutoff as two increasing connected linear functions (see Figure \ref{fig:model}):
 \begin{align}\label{pmodel}
    p(x)=\begin{cases}
        \pa{2}{l}x+\pb{2}{l}, & x<c-\pk{l}, \\
        \pa{1}{l}x+\pb{1}{l}, & c- \pk{l}\leq x<c, \\
        \pa{1}{r}x+\pb{1}{r}, & c\leq x\leq c+\pk{r},\\
        \pa{2}{r}x+\pb{2}{r}, & c+\pk{r}< x. \\
    \end{cases}
\end{align}
This model is restrictive enough to localize the cutoff, but at the same time flexible enough to fit the underlying data under Assumptions \ref{as5} and \ref{as6}, and to provide reliable estimates of  $j$. Moreover, since inference is local, we prefer to downweigh the data far away from the cutoff as much as possible. Having a second linear piece achieves this, relegating the data that is far away to a supporting role.

To ensure that the function $p$ is increasing, takes values between $0$ and $1$ and has discontinuity of size $j$ at $c$, where $j\geq \eta$, we solve a system of linear inequalities and equations (see Appendix \ref{App:ParamConst}). As a result, we impose the following restrictions on the values of the coefficients.

\begin{equation}\label{restriction}
\begin{aligned}[c]
\pa{2}{l}&\in \left[0,\frac{1-j}{c-\pk{l}-I_1}\right],\\
\pb{2}{l}&\in \left[-\pa{2}{l}I_1,1-j-\pa{2}{l}(c-\pk{l})\right],\\
\pa{1}{l}&\in \left[0,\frac{1-j-\pa{2}{l}(c-\pk{l})}{\pk{l}}\right],\\
\pb{1}{l}&=(c-\pk{l})(\pa{2}{l}-\pa{1}{l})+\pb{2}{l},
\end{aligned}
\quad
\begin{aligned}[c]
\pa{1}{r}&\in \left[0,\frac{1-\pa{1}{l}c-\pb{1}{l}-j}{\pk{r}}\right],\\
\pb{1}{r}&=(\pa{1}{l}-\pa{1}{r})c+\pb{1}{l}+j,\\
\pa{2}{r}&\in \left[0,\frac{1-\pb{1}{r}-(c+\pk{r}\pa{1}{r})}{I_2-c-\pk{r}}\right],\\
\pb{2}{r}&=(c+\pk{r})(\pa{1}{r}-\pa{2}{r})+\pb{1}{r}.
\end{aligned}
\\
\end{equation}
\\
It is important to notice that as we impose a positive lower bound on the jump size, the cutoff location is identifiable in our model.

We specify the prior distribution for the treatment probability function as follows. For now, consider parameters $j$, $c$, $\pk{l}$ and $\pk{r}$ as fixed. Since all coefficients are bounded, each function coefficient is assigned a conditional uniform prior over the interval indicated in \eqref{restriction}. For example, the prior on $\pb{2}{l}$ conditioned on $\pa{2}{l}$ and the fixed parameters is given by $\pi(\pb{2}{l}|\pa{2}{l})=\text{Uniform}(-\pa{2}{l}I_1,1-j-\pa{2}{l}(c-\pk{l}))$. Consequently, the joint conditional prior on the function coefficients is defined through the decomposition:
\begin{align*}
    \pi(\alpha_l,\beta_l,\alpha_r,\beta_r|c,j,\pk{r},\pk{l})=&\pi(\pa{2}{l}|j,k_{t})\pi(\pb{2}{l}|\pa{2}{l},c,j,\pk{r},\pk{l})\pi(\pa{1}{l}|\pb{2}{l},\pa{2}{l},c,j,\pk{r},\pk{l})\ldots\\
    &\ldots\pi(\pa{2}{r}|\pb{1}{r},\pa{1}{r},\alpha_l,\beta_l,c,j,\pk{r},\pk{l})\pi(\pb{2}{r}|\alpha_r,\pb{1}{r},\alpha_l,\beta_l,c,j,\pk{r},\pk{l}),
\end{align*}
where $\alpha_{\cdot}=(\pa{1}{\cdot},\pa{2}{\cdot})$, $\beta_{\cdot}=(\pb{1}{\cdot},\pb{2}{\cdot})$.

Next, we assume uniform priors on $\pk{l}$ and $\pk{r}$:
\begin{align*}
    \pi(\pk{l},\pk{r}|c)=\text{Uniform}(d_x,c-l_{25})\text{Uniform}(d_x,u_{25}-c),
\end{align*}
where $d_x$ is the upper quantile of the radii of balls centered at each $X_i$ containing exactly two other observed values of the score: one below $X_i$ and one above. If the score takes value on a grid, then $d_x$ is the distance between consecutive grid points. This lower bound ensures that there are multiple data points in the window $[c-\pk{l},c+\pk{r}]$ and therefore prevents convergence issues. For the same reason, we introduce the bounds $l_n$ and $u_n$. The value of $l_n$ ($u_n$) is defined as the minimum (maximum) value such that there are at least $n$ data points with scores below (above) $l_n$ ($u_n$). We set $n=25$ as the default.  The parameters $l_n$ and $u_n$ serve a purely technical role in stabilizing the MCMC algorithm. As long as $n$ remains relatively small, the results are unlikely to change significantly due to the intrinsic robustness of the linear part of the model.

Finally, we specify prior distributions for the main parameters of interest, namely $c$ and $j$. For the compliance rate, we assume it is difficult to predict beyond a lower bound. Therefore on $j$ we place a uniform prior between $\eta$ and $1$. On the other hand, the prior on the cutoff location should be guided by the expert knowledge. To start with, the support of the prior distribution should be narrowed down to an interval in which we expect the cutoff to be located, and that is separated from the boundary  points $I_1$ and $I_2$ to ensure stability of MCMC algorithm. If no further information is available, we assume uniform prior over that interval. Otherwise,  in the case of a continuous score, as default we consider a beta prior scaled and translated to that interval. If the score is discrete, then any prior weights can be assigned to the possible values within the interval, for instance through a beta-binomial distribution.

\subsection{Outcome model}\label{Section:outcome}
The next step is the estimation of the potential discontinuity in the outcome function $f$. Compared to the treatment probability function, this time we opt for a more flexible model. The outcome function does not play a primary role in the cutoff identification and has smaller influence on the stability of the treatment effect estimation. At the same time, we are interested in a point estimate at the cutoff. We need  reliable estimates of left and right limits at this point, and the goodness of fit far from it is not important per se. However, it is not an easy task to decrease the influence of distant points in a meaningful way. To illustrate this difficulty, let us consider that the underlying function is linear, then even points that are far from the cutoff carry information relevant to the point estimation at the cutoff. On the other hand, if the underlying function changes rapidly, the points far away are much less informative and can negatively influence the estimation at the cutoff. To tackle this challenge we propose the Local Trimmed Taylor Approximation (LoTTA) model (see Figure \ref{fig:model}): 
\begin{align}\label{fun_out}
    f(x)=\begin{cases}
        {\fl{0}}+{\fl{1}}(x-c), & \text{ for } \fk{l}<x<c,\\
        {\fl{0}}+{\fl{1}}(x-c)+\fl{2}(x-c)^2+\fl{3}(x-c)^3, & \text{ for } x\leq \fk{l}<c,\\
        {\fr{0}}+{\fr{1}}(x-c), & \text{ for } c\leq x<\fk{r},\\
        {\fr{0}}+{\fr{1}}(x-c)+\fr{2}(x-c)^2+\fr{3}(x-c)^3, & \text{ for } c<\fk{r}\leq x.\\
        \end{cases}
\end{align}
 We fit linear functions close to the cutoff, mimicking a local linear regression approach that has been successful in existing RD methods. Moreover, global high order polynomial approximations are not recommended in RDD, as they can produce poor estimates at the cutoff. \cite{Gelman2019} The purpose of the higher-order polynomial outside the window around $c$ is to minimize the effect of the data far from the cutoff on the treatment effect estimation. 
 
 The key element of our model is that the coefficients in the linear part are the same as in the cubic part that is fitted to the data further from the cutoff. This connection  helps to stabilize the inference at the cutoff. The big advantage of LoTTA is that it does not require manual adjustments. In particular, the window in which linear parts are fitted is given by model parameters $\fk{l}$ and $\fk{r}$. In the frequentist approach, the window selection is a crucial part of local linear regression; if the window bandwidth does not fulfill the theoretical requirements, significant undercoverage can be observed.\cite{Calonico2014} In our method  the window selection is directly incorporated into fitting the outcome model to the data. 

We use as a default the cubic function in the tail. For many applications it offers enough flexibility while keeping computing time relatively low. In Section \ref{diagnostic} we present diagnostics for the model fit. Naturally, the cubic function is not the only possibility. The most straightforward extension would be to increase the polynomial degree, however, we do not recommend this due to slower convergence and philosophical issues. \cite{Gelman2019} LoTTA can be extended to any function that has Taylor expansion at each point in the domain. Then the linear part is the first order Taylor approximation of the function in the tail. An example of such extension to inverse logit function is given in Appendix \ref{App1}.

It is important to point out that the function \eqref{fun_out} has discontinuity points not only at $c$ but also at $\fk{l}$ and $\fk{r}$. Such a modeling choice may not be intuitive at first as in a valid RDD we expect outcomes to follow a regular pattern on each side of the cutoff. Indeed, a big jump in the outcomes near the cutoff would be worrisome and could lead to skewed estimates of the treatment effect. However, the size of the discontinuity in \eqref{fun_out} is not arbitrary and diminishes as $\fk{l}$ and $\fk{r}$ get closer to $c$. One could think of similar modeling choices, for instance consider polynomial terms of the form $x-\fk{\cdot}$ instead of $x-c$. This would eliminate the discontinuities but at the same time change the focus from the point estimate at $c$ to the point estimate at $\fk{\cdot}$. The function is no longer expanded at $c$ - the crucial point in the treatment effect estimation, but at hyperparametres that are not well grounded in the data.   In practice, forced continuity at $\fk{l}$ and $\fk{R}$  can lead to convergence issues and consequently to a bad fit of the posterior outcome function. 

As indicated in Section \ref{sec1} the error term $\epsilon_x$ does depend on the score value. Precisely, we assume that $\epsilon_x\sim\mathcal{N}(0,\fs{r}{1})$ for $x\in [c,\fk{r}]$ and $\epsilon_x\sim\mathcal{N}(0,\fs{r}{2})$ for $x\in (\fk{r},I_2]$,  where $\fs{r}{2}\geq\fs{r}{1}$. Similarly, $\epsilon_x\sim\mathcal{N}(0,\fs{l}{1})$ for $x\in[\fk{l},c]$,  and $\epsilon_x\sim\mathcal{N}(0,\fs{l}{2})$ for $x\in[I_1,\fk{l})$, where $\fs{l}{2}\geq\fs{l}{1}$. Allowing the noise to increase further from the cutoff plays a double role. First, as we approach the boundary values of the score, it is reasonable to expect that the outcomes become more volatile. Second, the increased variance is a signal to the model that those points are of less importance, so the fit there can be compromised.

Now, as our model is fully defined, we consider the prior distributions. For the polynomial coefficients we set diffused hierarchical normal priors that approximate noninformative priors.
\begin{equation*}
\begin{aligned}[c]
&\fl{0}\sim \mathcal{N}(0,100),\\
&\fl{1}\sim \mathcal{N}(0,100),\\
&\fl{2}\sim \mathcal{N}(0,100(c-\fk{l})^{-0.5}),\\
&\fl{3}\sim \mathcal{N}(0,100(c-\fk{l})^{-0.5}),
\end{aligned}
\quad
\begin{aligned}[c]
&\fr{0}\sim \mathcal{N}(0,100),\\
&\fr{1}\sim \mathcal{N}(0,100),\\
&\fr{2}\sim \mathcal{N}(0,100(\fk{r}-c)^{-0.5}),\\
&\fr{3}\sim \mathcal{N}(0,100(\fk{r}-c)^{-0.5}).
\end{aligned}
\\
\end{equation*}
The priors on the quadratic and cubic terms are more dispersed as the linear part get reduced. We do so to take some mass from around $0$ and therefore favor longer linear part over longer cubic part with small nonlinear coefficients. If the outcome function is bounded, it can also be included in the coefficient priors. As the result, the posterior distribution of the treatment effect is only supported on the possible range of values (see Appendix \ref{App1}). 

We assume the following priors on the precisions of the error term:
\begin{equation*}
\begin{aligned}[c]
&\rho_1^R\sim \text{Gamma}(0.01,0.01),\\
&\rho_2^R\sim \text{Uniform}(0,\rho_1^R),
\end{aligned}
\quad
\begin{aligned}[c]
&\rho_1^L\sim \text{Gamma}(0.01,0.01),\\
&\rho_2^L\sim \text{Uniform}(0,\rho_1^R).
\end{aligned}
\\
\end{equation*}
We set noninformative priors on the main precision terms $\rho_1^R$ and $\rho_1^L$, which model the uncertainty of outcomes with corresponding scores within the interval $(\fk{l},\fk{r})$. As previously explained, we assume that variance in the tails may only increase. Consequently, the precision terms $\rho_2^R$ and $\rho_2^L$ are bounded between 0 and $\rho_1^R$ and $\rho_1^L$, respectively. 

Similarly as in the treatment model, we put uniform priors on $\fk{l}$ and $\fk{r}$:
\begin{equation*}
\begin{aligned}[c]
&\fk{l}\sim \text{Uniform}(l_{25},c-d_x),
\end{aligned}
\quad
\begin{aligned}[c]
&\fk{r}\sim \text{Uniform}(c+d_x,u_{25}).
\end{aligned}
\\
\end{equation*}
where $u_{25}$, $l_{25}$, and $d_x$ are defined in Subsection \ref{treatment_subsection}. As before, bounding domain of $\fk{l}$ and $\fk{r}$ is to stabilize the convergence of the MCMC algorithm. 

On final note: all the above priors are chosen for normalized datasets. Before applying the above model, scores should be divided by their range, and it should be ensured that values of $X$ and $Y$ lie on a similar scale.

\subsection{Combining the two models}\label{joinmodel}
Now that we have the two main ingredients, the treatment and the outcome model, they need to be combined. There are two main ways to do this: through joint estimation or a cut posterior. 
\cite{Jacob2017} Joint estimation combines the treatment and outcome models into a single joint likelihood. As a consequence, the posterior distributions of the parameters in these two models are dependent, as they share a common parameter $c$. The posterior distribution of the cutoff location is influenced both by the treatment allocation and the observed outcomes. In the cut posterior approach, the dependence is one-way: the treatment model is fitted first, and the resulting posterior samples of the cutoff location are plugged into the outcome model. Therefore, the posterior distribution of the cutoff location depends only on the treatment take-up data.

We observe in simulations (Appendix \ref{cut}) that a joint model produces superior results. This was expected, since the cutoff $c$ is a parameter in both the treatment and the outcome model and if there is a nonzero treatment effect, both the treatment and outcome data will contain information about $c$ (see also Figure \ref{fig:model}). A cut posterior only has access to one source of information about $c$ (the treatment data) while the joint posterior uses both sources. With a cut posterior, results may be heavily biased if a slight shift in the cutoff is detected based solely on the treatment data. A similar phenomenon is observed when a plug-in approach to cut-off estimation is taken (Appendix \ref{Plugin}).

The joint approach does require the treatment and outcome data to point to the same location of $c$. That is, if there is a discontinuity in the outcome function, it should be at the same location as the discontinuity in the treatment function. This is ensured by taking the standard RDD assumption that $\mathbb{E}[Y^{(t)} \mid X = x]$ is continuous everywhere \cite{Imbens2008} (a consequence of Assumption \ref{as3}), which prevents discontinuities in the outcome at a location other than $c$.  If there is no treatment effect, the cut posterior could be beneficial as no additional information on the cutoff location is contained in the outcome data. However, we did not observe negative effects of applying the joint model in this scenario; rather an inherent need of a stronger signal from the data about the jump in the treatment probability function. We move further discussion to Section \ref{simulations} and Appendix \ref{cut}.

The above considerations are linked to the ongoing discussion in Bayesian causal inference regarding the role of feedback between design and analysis stages.\cite{Li2023} However, this discussion usually centers on propensity score based models. Those advocating for fully separating the design stage from the analysis stage in these models typically argue based on philosophical considerations, such as that the outcome data should not contain any information about the treatment assignment process (represented by the propensity score) \cite{Rubin2008} and better empirical results if the feedback is eliminated, either through a cut posterior \cite{McCandle2010} or by plugging in the estimator of the propensity score into the outcome model. \cite{Zigler2013} Interestingly, in the case of RDD, both philosophical and pragmatic arguments are in favor of the joint model. 

\section{The Bayesian model as a validation and an estimation tool}\label{diagnostic}
We present a step-by-step approach to analyze datasets with an unknown cutoff. In the example data set, the cutoff is considered known. However, our analysis casts doubt on this assumption and illustrates how LoTTA can also be useful to validate a presumed known cutoff. As an example, we use data from Hlabsa HIV Care and Treatment Programme made available by Cattaneo et al.\cite{ART2023}  In this study, the score $X$ is the count of  CD4 cells per microliter of blood. The intervention is either immediate or deferred access to the antiretroviral therapy (ART). According to the national guidelines at the time, all patients with scores strictly below $350$ cells/$\upmu$l were eligible for the immediate start of the therapy. However, in the following analysis we show that a different value was likely used in practice.  We encode by $T=1$ the deferred start of the therapy and by $T=0$ the immediate start. In this way, we obtain the classic representation of an RDD, as introduced in Section~ \ref{setup}. The outcome is retention in care: $Y=1$ if there is any evidence of retention in care, $Y=0$ otherwise. In the following analysis, we fit the LoTTA model for binary outcomes to match the type of the data. The underlying principle is analogous to that of the LoTTA model for continuous outcomes presented in Section~\ref{Section:outcome}. The details for the binary LoTTA model are provided in Appendix \ref{App1}.

In this section, we discuss general steps of our Bayesian procedure to detect or validate a cutoff. In the next section, we give numerical results and additional details of this particular application. In Appendix \ref{chemo} we show how this framework can help in diagnosing a flawed RDD on the chemotherapy dataset previously analyzed by Cattaneo et al.\cite{Cattaneo2023}

\begin{figure}[ht]
     \centering
     \includegraphics[width=\textwidth]{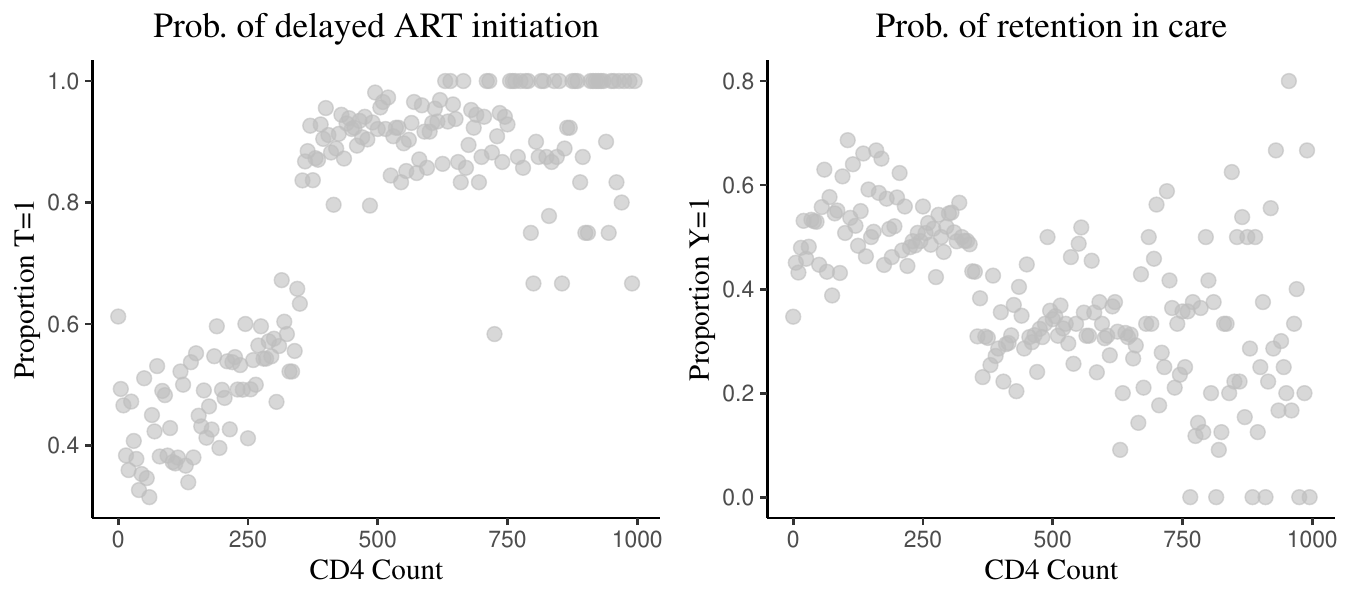}
     \caption{The binned ART data. We can observe that there is a visible decrease in the immediate access to ART and in the retention in care around $300$ and $400$ cells/$\upmu$l. The official guideline for the cutoff was 350, although as we argue in Section \ref{diagnostic} in practice it appears to have been 355.  }
     \label{ART:proportions}
 \end{figure}

\begin{figure}[ht]
     \centering
     \includegraphics[width=\textwidth]{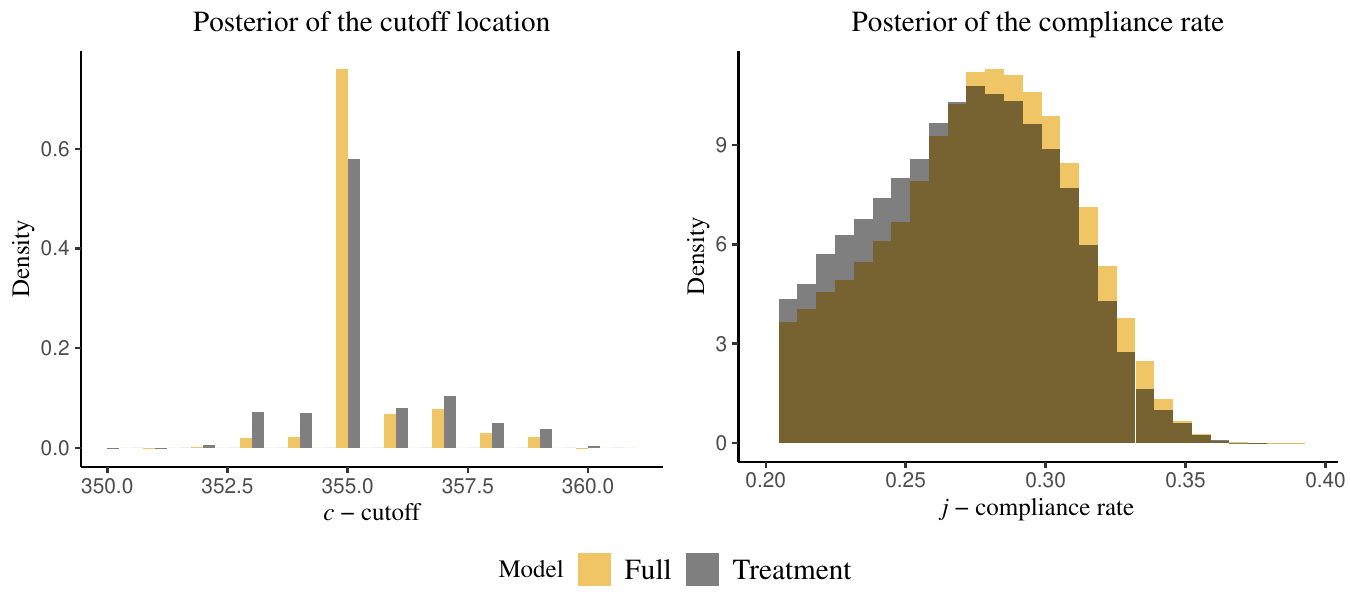}
     \caption{The posterior distributions of the cutoff location and compliance rate for the treatment and full LoTTA model. The posterior indicates a cutoff at $355$ instead of the cutoff at $350$ indicated by study guidelines. }
     \label{ART:histograms}
 \end{figure}

\begin{figure}[ht]
    \centering
    \includegraphics[width=0.88\textwidth]{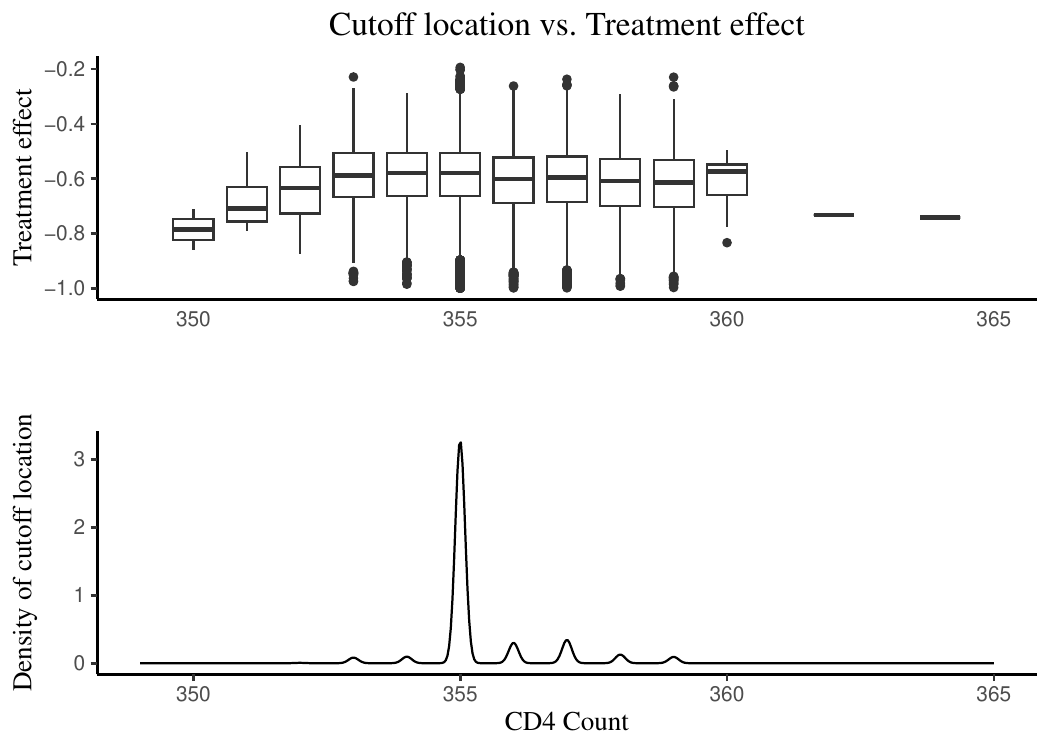}
    \caption{Visualization of the influence of the cutoff location on the estimated treatment effect in the ART application. The boxplot of the conditional treatment effect is paired with the density plot of the cutoff location to compare the contribution of each cutoff value in the final estimate of the treatment effect. } 
    \label{ART:joint}
\end{figure}

\begin{enumerate}[label=\textbf{\arabic*.}]
\item \textbf{Validate the model assumptions.}
    \begin{enumerate}[label=\textbf{1.\arabic*.}]
        \item \textbf{Plot the data.}  We start by plotting the binned treatment and outcome data (see Figure \ref{ART:proportions}). These plots give an idea of the cutoff location and the size of the potential treatment effect. Additionally, the plot of the treatment take-up indicates whether Assumptions \ref{as5} and \ref{as6}  are plausible. In particular, the plot should have at most two inflection points (due to a possible jump) and present an increasing trend; a decreasing trend requires recoding the data in order to use the model \eqref{pmodel} directly. 
        \item \textbf{Trim the data (optional).}\label{1.2} At this stage, data can be trimmed if there is some irregular behavior in the treatment take-up near the boundary points. If the score takes a wide range of values and the units close to the boundary points are edge cases, trimming might be necessary to ensure that assumptions \ref{as5} and \ref{as6} are justified. We trim ART data to only include score values between 50 and 950 cells/$\upmu$l as we notice less regular behavior and outliers near the boundary points. In the context of RDD, such modest trimming does not influence the reliability of the results. First, based on our prior knowledge and Figure \ref{ART:proportions} we expect the cutoff to be between 300 and 400 cells/$\upmu$l. Second, less than 200 cells/$\upmu$l is an indication for AIDS diagnosis, and values between 500 and 1000 cells/$\upmu$l are within the normal range.\cite{Garcia2024} Therefore, we are not discarding informative data points regarding the treatment effect at the cutoff.
    \end{enumerate}
  \item \textbf{Validate the existence of a cutoff.}
  \begin{enumerate}[label=\textbf{2.\arabic*.}]
      
        \item \textbf{Fit the treatment model.} We fit the treatment model \eqref{pmodel} separately of the outcome model; this is to detect potential issues within the analysis. If the treatment and the full model lead to different conclusions, it may signal either anomalies in the dataset, a small compliance rate with respect to the amount of data, or misfit of the model. In the ART application we set a uniform prior on the cutoff location between $300$ and $400$ cells/$\upmu$l, and bound the jump size from below by the default value $0.2$. 
        \item \textbf{Fit the full model.}
        Next, we fit the full model. In this application, we use the LoTTA model for binary outcomes. The specification of this model can be found in Appendix \ref{App1}. 
        \item \textbf{Plot the histograms.}\label{2.3} We start by comparing the histograms of the posterior distributions of $c$ and $j$ in the treatment and full model.   Figure \ref{ART:histograms} shows that the posterior distributions of the cutoff location are similar and concentrate on the same value. Notably, the joint model assigns even higher mass to $355$. This is what we expect if the design is valid, since there appears  to be a nonzero treatment effect. Similarly, on Figure \ref{ART:histograms} we observe that the histograms of the jump size also largely overlap. Importantly, the distribution of $j$ is not skewed towards the lower bound. The left tail is thick, but the mode appears to be around $0.27$. Finally, we check that the posterior distribution of the treatment effect is unimodal. If it is not the case, this may point to multiple cutoffs being identified by the model, or it may signal problems with the model fit. 
  \end{enumerate}
  \paragraph{Intermezzo: proceeding after finding the cutoff in a different location.} In this particular study the cutoff was known; if we were validating the design at the given cutoff value  we should conclude that this RDD is not valid. At the same time, the data shows evidence that a different cutoff value was used. Whether one should continue the analysis, allowing for the shifted cutoff, requires careful consideration in each case, and if possible contacting the entity that conducted the experiment. To communicate the results to collaborators with limited statistical background, we recommend using graphical representations of the posterior distribution - such as histograms of the cutoff location and compliance rate (see point \ref{2.3}), as well as the plot of the posterior treatment probability function (see point \ref{3.1}).  In some cases the treatment effect may still be estimated if we believe that a significant fraction of decision makers used consistently the same cutoff (see Appendix \ref{App:iden_res}). For instance, when the cutoff shift is thought to result from rounding the scores.  It is important to keep in mind that consulting decision makers may help us understand why a cutoff shift might have occurred; however, it does not necessarily allow us to determine the cutoff location with full certainty.  For this reason, if the decision is made to proceed with the analysis, we recommend reporting estimates from the Bayesian model rather than relying on plug-in methods, which can lead to highly unreliable results (see Appendix \ref{Plugin}).  One should also consider whether shifting the cutoff could have influenced the results. Precisely, whether the intention-to-treat effect would be the same had the cutoff been set at the shifted value at the beginning of the experiment and the guidelines had been respected. As an example, this may be violated if breaking the rules causes doctors to pay more attention to their patients. 
  
  In our case, the posterior mass concentrates strongly at one value, hence it is plausible that the majority of the doctors used the same cutoff point. Moreover, the shift of the cutoff point is relatively small. However, we do not have any additional information on the implementation of the guidelines that could explain the cutoff shift.  To illustrate the use of LoTTA, we continue the analysis of the ART dataset.
   \item \textbf{Check the model fit and the sensitivity to the model parameters.}
   \begin{enumerate}[label=\textbf{3.\arabic*.}]
        \item \textbf{Plot the posterior functions.}\label{3.1} Each sample from the posterior gives us a treatment probability function and an outcome function. In Figure \ref{ART:plots}, we plot the pointwise median values along with the pointwise $95\%$ credible intervals of those functions. Underneath them we plot the binned data. We bin the data separately on both sides of the MAP cutoff estimate to visualize the potential jumps.
         \item \textbf{Check the stability of the results (optional).} If there is a meaningful focus region in the data, then the model can be fitted to it by trimming the data at this region's boundaries. Then we repeat steps~\ref{2.3} and \ref{3.1}, and compare the results to the global fit. In ART application we argue that such region is between 200 cells/$\upmu$l and 500 cells/$\upmu$l (see Step~\ref{1.2} ). Figure \ref{ART:plots} shows the stability of our results. 
         \item \textbf{Check the sensitivity with respect to $\eta$.} If the posterior distribution of the compliance rate places substantial mass near the lower bound, the analysis should be repeated with a smaller value of $\eta$ to assess the stability of the results. However, the alternative value of $\eta$ should not be set too close to zero.  In Appendix \ref{APP:ART_sens}, we present the results of a sensitivity analysis on the ART dataset using $\eta = 0.05$.
         \item \textbf{Check the sensitivity to the prior on the cutoff location.} In the main analysis, we assumed a uniform prior on the cutoff location. In the ART dataset, the cutoff was presumably known; therefore, we assess the stability of the results under a strongly informative prior (see Appendix \ref{APP:ART_sens}). Conversely, if an informative prior was used in the main analysis, a non-informative prior should also be investigated. 
        
    \end{enumerate}  
    \item \textbf{Report estimates and credible regions.} We report the MAP estimates of the treatment effect, the cutoff $c$, and the compliance rate $j$ with their corresponding $95\%$ highest density intervals.  
\end{enumerate}
\paragraph{Interpretation of the results.} The posterior distribution of the treatment effect is marginalized in particular over cutoff location, which we treat as a random variable. The resulting  MAP estimate of the treatment effect is the most likely value of the treatment effect across plausible values of $c$. This means that  different cutoff values contribute to the final estimate of the treatment effect. What is important is that for each vector of parameters, the treatment effect is identifiable. To better understand the influence of the cutoff location on the treatment effect we recommend plotting the joint distribution of $c$ and $\tau$ (see Figure \ref{ART:joint}). Ideally, we would like to observe similar estimates of the treatment effect across the plausible values of $c$.
\begin{figure}[ht]
    \centering
    \includegraphics[width=0.8\textwidth]{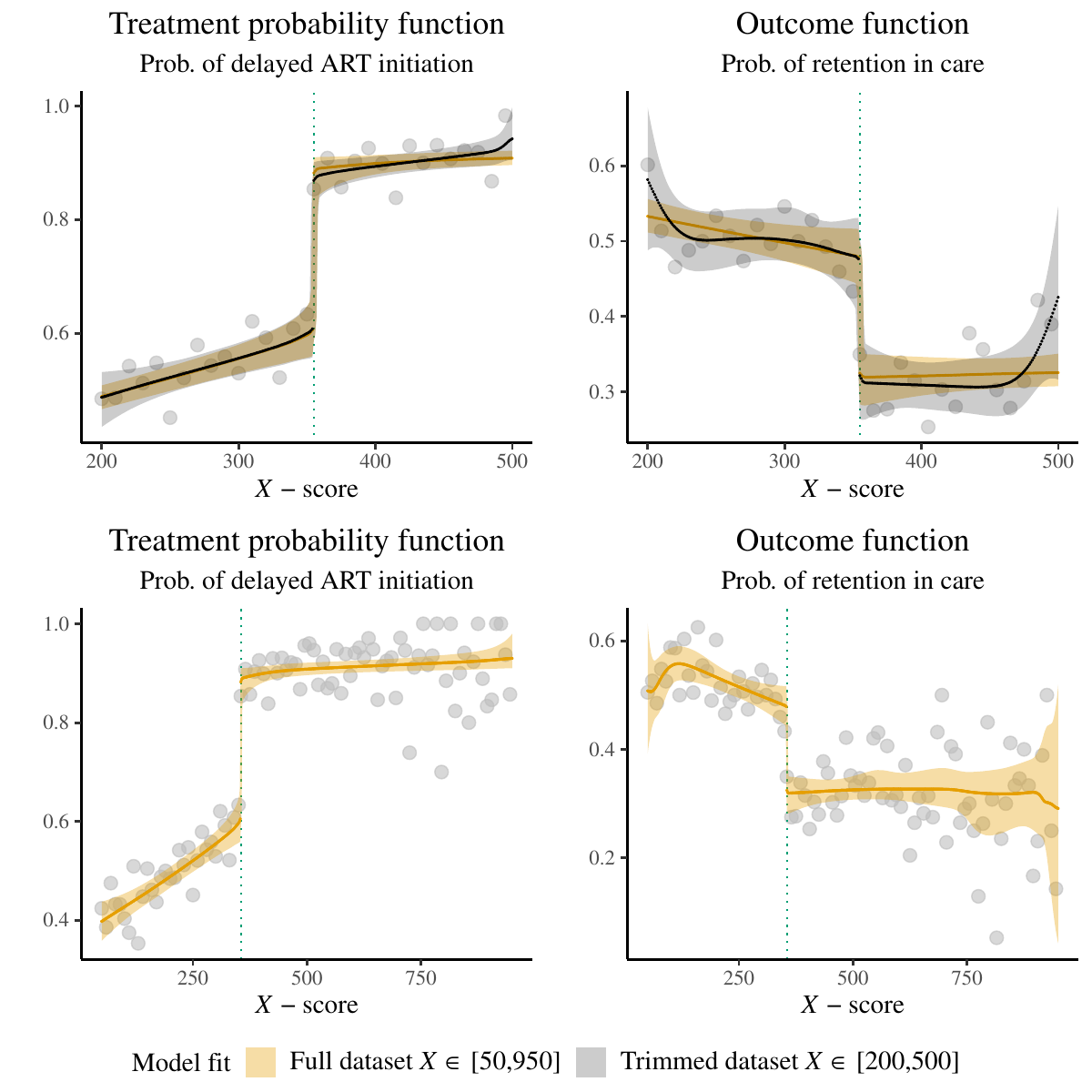}
    \caption{Median posterior treatment and outcome functions with 95\% credible band for ART data. In the upper row we plot the global fit of the LoTTA model. In the bottom row we compare the global fit with the local fit on trimmed data. }
    \label{ART:plots}
\end{figure}

\paragraph{Further validation of RDD with unknown cutoff.}
In the preceding part, we focused solely on the cutoff validation, but it is a good practice to check for the covariate balance and score manipulation as well.  Cattaneo et al.\cite{Cattaneo_Idrobo_Titiunik_2020} provide a summary of the additional falsification methods. As they all require a fixed cutoff point, we recommend to set it to the MAP estimate of $c$ from the full model and then perform further falsification tests. If there is some anomaly in the region around the MAP estimate, we should be able to detect it without checking multiple points.

\subsection{(Un)known cutoff - ART application}
 The analysis described in the previous subsection revealed that the official guideline for the immediate ART initiation was not followed by medical practitioners, and that a different cutoff criterion was likely used in practice. 
  The official cutoff of $350$ cells/$\upmu$l appears to be different from the empirical cutoff of $355$ ~cells/$\upmu$l at which the discontinuity in the treatment take-up most probably occurs. This disparity had so far gone unnoticed, even though the data was previously analyzed according to the established best practices, including graphical inspection.  \cite{Bor2017}$^,$\cite{Cattaneo2023} While the difference between 355 and 350 may seem not so big, this difference is crucial for correct estimation of the treatment effect.  Based on the results in Table \ref{table1} for LLR at $c=350$ in the full and trimmed datasets, we speculate that the close proximity of the two points caused an overestimation of the compliance rate at $350$ cells/$\upmu$l. Consequently, this overestimation led to a validation of the design at an incorrect point.  This example demonstrates the need of additional tools that do not aim at falsifying RDD assumptions at a single, given point but take into consideration a broader window of data. This means removing the hidden assumption that a design is either valid at the cutoff given by official guidelines or no cutoff was used at all. 

  In Table \ref{table1}, we compare the results obtained through LoTTA and LLR. All the results for LLR in this and in the following sections were generated using \texttt{rdrobust} package in \texttt{R}.\cite{Calonico2014}$^,$\cite{Calonico2017} In the case of LLR, we estimate the treatment effect and the compliance rate $j$ both at the given cutoff and at LoTTA's MAP estimate. We consider two datasets: the (almost) full one that includes scores between $50$ and $950$ cells/$\upmu$l, and the trimmed one that includes scores between $200$ and $500$ cells/$\upmu$l. In the full dataset, LLR computed at $350$ cells/$\upmu$/l gives wider confidence intervals and a smaller estimate of the compliance rate, but the results are still fairly similar. The biggest difference occurs for the trimmed dataset. The estimates of LLR both at $355$ cells/$\upmu$l and particularly at $350$ cells/$\upmu$l become  unstable, visible from the widening of the confidence intervals. However, the problem may not only lie in the cutoff point but also in the nature of the data. Indeed, the trimmed dataset contains 3507 data points, while the full data set contains 6819 data points, and the outcome variable is binary; LLR is not well adjusted to this type of outcomes. On the other hand, LoTTA can be easily modified to deal with binary outcomes (see Appendix \ref{App1}), which in turn leads to more stable results. The additional advantage is also achieved through the lower bound on the compliance rate.

\begin{table}[ht]
\centering
\resizebox{0.8\textwidth}{!}{
\begin{tabular}{|lccc|}

\rowcolor[HTML]{EFEFEF} 
\hline
\multicolumn{4}{|l|}{Full dataset: $X \in {[}50,950{]}$}                                                                                                              \\ \hline
\multicolumn{1}{|l|}{}                               & \multicolumn{2}{c|}{LLR}                                                            & LoTTA                 \\ 
\multicolumn{1}{|l|}{}                               & \multicolumn{1}{c}{c=350}               & \multicolumn{1}{c|}{c=355}               & \multicolumn{1}{l|}{} \\ \cline{2-4} 
\multicolumn{1}{|l|}{LATE} & \multicolumn{1}{c}{-0.64 (-0.99,-0.29)} & \multicolumn{1}{c|}{-0.55 (-0.85,-0.24)} & -0.56 (-0.85,-0.38)   \\ 
\multicolumn{1}{|l|}{Compliance rate}                & \multicolumn{1}{c}{0.19 (0.11,0.28)}    & \multicolumn{1}{c|}{0.26 (0.17,0.34)}    & 0.28 (0.2,0.33)       \\ 
\multicolumn{1}{|l|}{Cutoff location}                & \multicolumn{1}{c}{--}                   & \multicolumn{1}{c|}{--}                   & 355 (354,358)         \\ 

\rowcolor[HTML]{EFEFEF} 
\hline
\multicolumn{4}{|l|}{Trimmed dataset: $X \in {[}200,500{]}$}                                                                                                          \\ \hline
\multicolumn{1}{|l|}{}                               & \multicolumn{2}{c|}{LLR}                                                            & LoTTA                 \\ 
\multicolumn{1}{|l|}{}                               & \multicolumn{1}{c}{c=350}               & \multicolumn{1}{c|}{c=355}               & \multicolumn{1}{l|}{} \\ \cline{2-4} 
\multicolumn{1}{|l|}{LATE} & \multicolumn{1}{c|}{-0.47 (-1.43,0.48)}  & \multicolumn{1}{c|}{-0.47 (-1.12,0.17)} & -0.63 (-0.97,-0.31)   \\  
\multicolumn{1}{|l|}{Compliance rate}                & \multicolumn{1}{c|}{0.09 (-0.05,0.22)}   & \multicolumn{1}{c|}{0.19 (0.05,0.33)}    & 0.27 (0.2,0.32)       \\ 
\multicolumn{1}{|l|}{Cutoff location}                & \multicolumn{1}{c|}{--}                   & \multicolumn{1}{c|}{--}                   & 355 (354,358)         \\ \hline
\end{tabular}
}
\caption{ Reanalysis of ART data. LoTTA refers to the LoTTA model for binary outcomes. For the LLR robust 95\% confidence intervals are given in the parentheses. For LoTTA, MAP estimates are given along with 95\% highest density intervals.}
\label{table1}
\end{table}
\paragraph{Accounting for heterogeneity in cutoff rules.} In the case of a shifted cutoff, a natural question arises: is the treatment effect identifiable, given that multiple cutoff rules may have been used? In the ART example, while the data suggest that most physicians followed a cutoff of 355, it is plausible that a small fraction adhered to the official cutoff of 350. In the Appendix \ref{App:iden_res}, we argue that in such cases, the causal estimand of interest is $$\tau_c=\mathbb{E}[Y^{(1,c)}-Y^{(0,c)}|X=c,C_T=C,R=c],$$ where the potential outcomes are indexed by both the treatment received $t\in\{0,1\}$ and the cutoff rule 
$c$ employed by a decision maker. This formulation allows for heterogeneity across decision makers based on the cutoff rule they employed. For instance, in the ART example, physicians using the shifted cutoff may have worked at different clinics than those using the official cutoff, or may have paid closer attention to patients due to consciously bending the rules. We show that the treatment effect $\tau_c$ is identifiable under mild conditions implying that neither scores nor choice of a decision maker can be systematically manipulated, and continuity assumptions analogous to those in a standard fuzzy RDD. Specifically, $\tau_c$ is identifiable through the same formula as $\tau$ (see Equation \eqref{tau}). However, the interpretation is more nuanced than in the classical fuzzy RDD. Specifically, the average treatment effect $\tau_c$ applies to compliers with scores equal to $c$, whose treatment assignment followed the decision rule associated with cutoff $c$. Whether the resulting estimate can be interpreted as the treatment effect for the broader population of compliers with score $c$ depends on the context in which the shifted cutoff arises, as discussed further in Appendix \ref{App:iden_res}.

\section{Simulation results}\label{simulations}
We perform a series of simulation studies to validate our method. We consider three main scenarios: sharp design with a known cutoff (Scenario 1), and fuzzy design with unknown cutoff and two varying jump sizes in the treatment probability function (Scenarios 2 and 3). We compare our results to local linear regression with robust confidence intervals from \texttt{rdrobust} package.\cite{Calonico2014} The sharp design is the clearest setup for the comparison as all the methods have access to the cutoff location. For this reason, we include it in our study beside the fuzzy design. The comparison in the fuzzy design is harder to interpret because there is no other available method that includes cutoff detection or that bounds the discontinuity in the treatment assignment away from $0$. In Scenarios 2 and 3, local linear regression, which uses the prespecified cutoff location while LoTTA does not, is therefore best viewed as an oracle benchmark to indicate to what extent extra uncertainty is added by having an unknown cutoff. Additionally, we investigate the effect of the feedback between the outcome and the treatment model on the estimation of the cutoff location. Specifically, we compare the performance of the joint model and the treatment-only model in estimating the cutoff location.

\begin{figure}
    \centering
    \includegraphics[width=\textwidth]{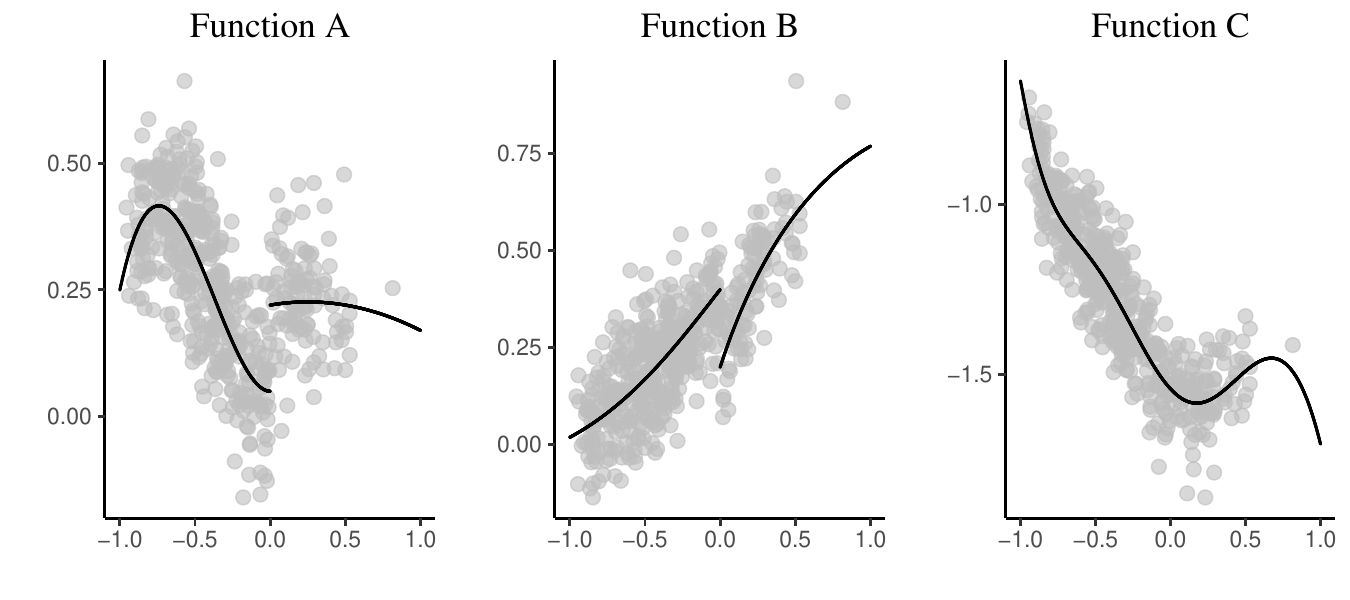}
    \caption{Outcome functions used in simulations: cubic, non-polynomial, no treatment effect.}
    \label{fig:sim_fun}
\end{figure}

In each simulation setup, we generate 1000 datasets consisting of 500 data points. Based on simulations previously considered in RDD literature, \cite{Calonico2014}$^,$\cite{Branson2019} we sample score values according to $X\sim 2 Z -1$, where $Z\sim beta(2,4)$, and we set the cutoff point at $0$. Therefore the majority of the data points is located to the left of the cutoff. We set uniform prior on $c$ between $-0.8$ and $0.2$; the interval $[-0.8,0.2]$ contains around $80\%$ to $85\%$ of all score samples. The outcomes were sampled according to ${y_i\sim\mu_j(x_i)+\epsilon}$, with different mean functions and $\epsilon\sim \mathcal{N}(0,0.1)$. We denote by $g(x)=(1+\exp{(-x)})^{-1}$ the inverse logit function and we consider the following functions (see Figure \ref{fig:sim_fun}).
\begin{enumerate}
    \item[A.] Cubic: $\mu_A(x)=\mathbbm{1}(x<c)[1.8x^3+2x^2+0.05]+\mathbbm{1}(x\geq c)[0.05x-0.1x^2+0.22]$.
    \item[B.] Non-polynomial: $\mu_B(x)= \mathbbm{1}(x<c)[g(2x)-0.1]+\mathbbm{1}(x\geq c)[0.6(\ln{(2x+1)}-0.15x^2)+0.20]$.
    \item[C.] No treatment effect: $\mu_C(x)=
    -0.952- 0.27x+ 0.118x^2 +0.121x^3 + 0.254x^4 - 0.3x^5-0.19x^6-0.5g(10(x+1))+\sin(5x-2).
$
\end{enumerate} 
The first function is a polynomial without a linear part on the left hand side to add difficulty for our model. The second function has an infinite Taylor expansion but a regular shape that we find realistic to occur in empirical datasets. Finally, the last function also admits infinite Taylor expansion and corresponds to the scenario of no treatment effect as it is continuous and differentiable in the whole domain. The discontinuity at the cutoff equals to $0.17$, $-0.2$ and $0$ for Scenarios A, B and C, respectively. We present additional results in Appendix \ref{Lee} that include two regularly used functions in the RDD literature, called `Lee' and `Ludwig', which demonstrate that LoTTA is more robust than the global cubic polynomial. 

The treatment allocation $t_i$ was sampled according to $ber(p_j(x_i))$. For the treatment probability function, we considered two increasing functions, with a jumpsize of $0.55$ and $0.3$.
\begin{enumerate}
    \item $p_1(x)=\mathbbm{1}(x<c)[(x+1)^4/15+0.05]+\mathbbm{1}(x\geq c)[g((8.5x-1.5))/10.5-g(-1.5)/10.5+1/15+0.65]$.
    \item $p_2(x)=\mathbbm{1}(x<c)[(x+1)^4/15+0.05]+\mathbbm{1}(x\geq c)[g((8.5x-1.5))/10.5-g(-1.5)/10.5+1/15+0.35]$.
\end{enumerate}
The jump size of $0.3$ is particularly challenging considering that the treatment data is binary and sparse on the right side of the cutoff. 

We summarize the simulation results in Table \ref{Tabres1} and Table \ref{Tabres23}, where we compare LoTTA with LLR, and in Table \ref{Tabres4}, where we compare the joint model with the treatment-only model in estimating the cutoff location. RMSE and AE refer to root mean squared error and absolute error of the point estimates. 

In the case of the sharp designs, LoTTA and LLR give competitive results. LoTTA tends to have narrower credible intervals, while offering similar coverage. It is in the fuzzy design that the differences become more apparent. LLR presents less stable behavior, which becomes particularly noticeable for the low compliance rate of $0.3$. Unlike LoTTA, LLR does not impose any additional structure on the treatment probability function to decrease the volatility of its fit. As the mean values of the statistics are heavily influenced by outliers, in the tables we include the median interval length, and the median absolute error and bias for an alternative comparison of the two methods. 

First, we focus on the larger jump size. In Scenarios 2A and 2B, LoTTA results in shorter intervals with better coverage compared to LLR. Although the cutoff is unknown it can be precisely estimated (see Table \ref{Tabres4}). The discontinuities in functions A and B help in localizing the cutoff, as estimates from the joint model are more precise than those from the treatment-only model. Moreover, the structure imposed on the treatment probability function effectively reduces the uncertainty of the LATE estimates. Consequently, LoTTA more accurately identifies the sign of the treatment effects. In Scenario 2C,  median LoTTA credible interval is wider than median LLR confidence interval. Moreover, median credible interval of the cutoff location is wider compared to scenario 2A and 2B. This aligns with our intuition: the outcome function does not provide information about the cutoff location, leading to less precise estimates of both the cutoff and the treatment effect. Finally, we do not expect the results to change substantially for smaller values of $\eta$, since the true jump size is considerably larger than the lower bound $\eta = 0.2$.

In the fuzzy design with a low compliance rate, the instability of LLR becomes more pronounced. In Scenarios 3A and 3B, median confidence intervals are relatively long, while the coverage is significantly below the nominal value. The additional structure of the probability function in LoTTA model becomes particularly advantageous, as it narrows credible intervals while ensuring good coverage. In Scenario 3, we expect that lowering $\eta$ would lead to wider credible intervals and less precise cutoff estimation particularly in Scenario 3C.

It is worth noticing that in the above scenarios we do not observe any negative effects of the feedback between treatment and outcome models. When a treatment effect is present, feedback from the outcome model improves the estimation of the cutoff location; this is particularly visible in Scenarios 3A and 3B. In the absence of a treatment effect, although cutoff estimation remains influenced by the outcome model fit, the final results remain largely consistent with those from the treatment-only model.

\begin{table}

\centering

\begin{tabular}{|lcccccc|}
\hline
\rowcolor[HTML]{EFEFEF} 
\multicolumn{7}{|l|}{Scenario 1: Sharp design} \\ \hline
\multicolumn{1}{|l|}{} &
  \multicolumn{2}{c|}{Scenario 1A } &
  \multicolumn{2}{c|}{Scenario 1B } &
  \multicolumn{2}{c|}{Scenario 1C } \\
  \multicolumn{1}{|l|}{} &
  \multicolumn{2}{c|}{(cubic, $\tau=0.17$)} &
  \multicolumn{2}{c|}{(misspecified, $\tau=-0.2$)} &
  \multicolumn{2}{c|}{(misspecified, $\tau=0$)} \\
\multicolumn{1}{|l|}{} &
  \begin{tabular}[c]{@{}c@{}}LoTTA\\ Known \\Cutoff\end{tabular} &
  \multicolumn{1}{c|}{\begin{tabular}[c]{@{}c@{}}LLR\\ Known \\Cutoff\end{tabular}} &
  \begin{tabular}[c]{@{}c@{}}LoTTA\\ Known \\Cutoff\end{tabular} &
  \multicolumn{1}{c|}{\begin{tabular}[c]{@{}c@{}}LLR\\ Known \\Cutoff\end{tabular}} &
  \begin{tabular}[c]{@{}c@{}}LoTTA\\ Known \\Cutoff\end{tabular} &
  \begin{tabular}[c]{@{}c@{}}LLR\\ Known \\Cutoff\end{tabular} \\ \cline{2-7} 
\rowcolor[HTML]{EFEFEF} \multicolumn{1}{|l|}{\textbf{LATE:} RMSE} &
  0.04 &
  \multicolumn{1}{c|}{0.05} &
  0.04 &
  \multicolumn{1}{c|}{0.05} &
  0.04 &
  0.05 \\
   \multicolumn{1}{|l|}{Mean Bias} &
  0.01 &
  \multicolumn{1}{c|}{0.00} &
  0.00 &
  \multicolumn{1}{c|}{-0.00} &
  0.01 &
  0.00 \\
\rowcolor[HTML]{EFEFEF} \multicolumn{1}{|l|}{Average CI$^*$ Length} &
  0.15 &
  \multicolumn{1}{c|}{0.19} &
  0.13 &
  \multicolumn{1}{c|}{0.19} &
  0.13 &
  0.19 \\
 \multicolumn{1}{|l|}{Empirical Coverage} &
  0.92 &
  \multicolumn{1}{c|}{0.93} &
  0.94 &
  \multicolumn{1}{c|}{0.92} &
  0.94 &
  0.92 \\
\rowcolor[HTML]{EFEFEF}\multicolumn{1}{|l|}{\begin{tabular}[c]{@{}l@{}}Correct Sign$^{**}$ \end{tabular}} &
  0.99 &
  \multicolumn{1}{c|}{0.91} &
  0.99 &
  \multicolumn{1}{c|}{0.96} &
  -- &
  -- \\
 \hline
\end{tabular}%

\caption{Simulation results for the sharp design. Both LoTTA and LLR have access to the cutoff location.$^{*}$`CI' stands for $95\%$ credible interval or confidence interval depending on the context, $^{**}$`Correct sign' refers to the proportion of intervals identifying the sign of the treatment effect.}\label{Tabres1}
\end{table}
\begin{table}

\resizebox{\textwidth}{!}{
\begin{tabular}{|lcccccc|}
\hline
\rowcolor[HTML]{EFEFEF} 
\multicolumn{7}{|l|}{Scenario 2: Fuzzy design, $j=0.55$} \\ \hline
\multicolumn{1}{|l|}{} &
  \multicolumn{2}{c|}{Scenario 2A } &
  \multicolumn{2}{c|}{Scenario 2B } &
  \multicolumn{2}{c|}{Scenario 2C } \\
  \multicolumn{1}{|l|}{} &
  \multicolumn{2}{c|}{(cubic, $\tau=0.31$)} &
  \multicolumn{2}{c|}{(misspecified, $\tau=-0.36$)} &
  \multicolumn{2}{c|}{(misspecified, $\tau=0$)} \\
\multicolumn{1}{|l|}{} &
  \begin{tabular}[c]{@{}c@{}}LoTTA\\ Unknown\\ Cutoff\end{tabular} &
  \multicolumn{1}{c|}{\begin{tabular}[c]{@{}c@{}}LLR\\ Known\\ Cutoff\end{tabular}} &
  \begin{tabular}[c]{@{}c@{}}LoTTA\\ Unknown\\ Cutoff\end{tabular} &
  \multicolumn{1}{c|}{\begin{tabular}[c]{@{}c@{}}LLR\\ Known\\ Cutoff\end{tabular}} &
  \begin{tabular}[c]{@{}c@{}}LoTTA\\ Unknown\\ Cutoff\end{tabular} &
  \begin{tabular}[c]{@{}c@{}}LLR\\ Known\\ Cutoff\end{tabular} \\
  \cline{2-7} 
\rowcolor[HTML]{EFEFEF} \multicolumn{1}{|l|}{\textbf{LATE:} RMSE } &
  0.1 &
  \multicolumn{1}{c|}{9.87} &
  0.08 &
  \multicolumn{1}{c|}{0.67} &
  0.08 &
  0.13 \\
 \multicolumn{1}{|l|}{Mean Bias} &
  0.04 &
  \multicolumn{1}{c|}{0.47} &
  -0.01 &
  \multicolumn{1}{c|}{-0.1} &
  0.02 &
  0.00 \\
   \rowcolor[HTML]{EFEFEF}\multicolumn{1}{|l|}{Median Bias} &
  0.04 &
  \multicolumn{1}{c|}{0.01} &
  0.00 &
  \multicolumn{1}{c|}{0.01} &
  0.01 &
  0.00 \\
\multicolumn{1}{|l|}{Median AE} &
  0.06 &
  \multicolumn{1}{c|}{0.09} &
  0.05 &
  \multicolumn{1}{c|}{0.1} &
  0.04 &
  0.06 \\
\rowcolor[HTML]{EFEFEF} \multicolumn{1}{|l|}{Median CI$^*$ Length} &
  0.54 &
  \multicolumn{1}{c|}{0.54} &
  0.52 &
  \multicolumn{1}{c|}{0.55} &
  0.35 &
  0.32 \\
\multicolumn{1}{|l|}{Mean CI$^*$ Length} &
  0.57 &
  \multicolumn{1}{c|}{4.66} &
  0.54 &
  \multicolumn{1}{c|}{1.12} &
  0.38 &
  0.41 \\
\rowcolor[HTML]{EFEFEF} \multicolumn{1}{|l|}{Empirical Coverage} &
  0.97 &
  \multicolumn{1}{c|}{0.93} &
  0.98 &
  \multicolumn{1}{c|}{0.92} &
  0.96 &
  0.96 \\
\multicolumn{1}{|l|}{\begin{tabular}[c]{@{}l@{}}Correct Sign$^{**}$\end{tabular}} &
  0.94 &
  \multicolumn{1}{c|}{0.65} &
  0.96 &
  \multicolumn{1}{c|}{0.75} &
  -- &
  -- \\
\rowcolor[HTML]{EFEFEF} \multicolumn{1}{|l|}{\textbf{Cutoff:} RMSE} &
  0.01 &
  \multicolumn{1}{c|}{--} &
  0.01 &
  \multicolumn{1}{c|}{--} &
  0.02 &
  -- \\
\multicolumn{1}{|l|}{\textbf{Compliance:} RMSE} &
  0.1 &
  \multicolumn{1}{c|}{0.2} &
  0.09 &
  \multicolumn{1}{c|}{0.19} &
  0.1 &
  0.19 \\ \hline

\rowcolor[HTML]{EFEFEF} 
\multicolumn{7}{|l|}{Scenario 3: Fuzzy design, $j=0.3$} \\ \hline
\multicolumn{1}{|l|}{} &
  \multicolumn{2}{c|}{Scenario 3A } &
  \multicolumn{2}{c|}{Scenario 3B } &
  \multicolumn{2}{c|}{Scenario 3C } \\
  \multicolumn{1}{|l|}{} &
  \multicolumn{2}{c|}{(cubic, $\tau=0.57$)} &
  \multicolumn{2}{c|}{(misspecified, $\tau=-0.67$)} &
  \multicolumn{2}{c|}{(misspecified, $\tau=0$)} \\
\multicolumn{1}{|l|}{} &
  \begin{tabular}[c]{@{}c@{}}LoTTA\\ Unknown\\ Cutoff\end{tabular} &
  \multicolumn{1}{c|}{\begin{tabular}[c]{@{}c@{}}LLR\\ Known \\ Cutoff\end{tabular}} &
  \begin{tabular}[c]{@{}c@{}}LoTTA\\ Unknown\\ Cutoff\end{tabular} &
  \multicolumn{1}{c|}{\begin{tabular}[c]{@{}c@{}}LLR\\ Known \\ Cutoff \end{tabular}} &
  \begin{tabular}[c]{@{}c@{}}LoTTA\\ Unknown\\ Cutoff\end{tabular} &
  \begin{tabular}[c]{@{}c@{}}LLR\\ Known \\ Cutoff\end{tabular} \\ \cline{2-7} 
\rowcolor[HTML]{EFEFEF} \multicolumn{1}{|l|}{\textbf{LATE:} RMSE } &
  0.19 &
  \multicolumn{1}{c|}{241.48} &
  0.18 &
  \multicolumn{1}{c|}{15518.54} &
  0.15 &
  1227.65 \\
  \multicolumn{1}{|l|}{Mean Bias} &
  0.02 &
  \multicolumn{1}{c|}{9.61} &
  0.03 &
  \multicolumn{1}{c|}{-495.1} &
  0.06 &
  39.52 \\
   \rowcolor[HTML]{EFEFEF}\multicolumn{1}{|l|}{Median Bias} &
  0.02 &
  \multicolumn{1}{c|}{-0.02} &
  0.04 &
  \multicolumn{1}{c|}{0.07} &
  0.05 &
  0.01 \\
\multicolumn{1}{|l|}{Median AE } &
  0.11 &
  \multicolumn{1}{c|}{0.24} &
  0.11 &
  \multicolumn{1}{c|}{0.26} &
  0.09 &
  0.11 \\
\rowcolor[HTML]{EFEFEF} \multicolumn{1}{|l|}{Median CI$^*$ Length} &
  0.77 &
  \multicolumn{1}{c|}{1.49} &
  0.73 &
  \multicolumn{1}{c|}{1.48} &
  0.63 &
  0.65 \\
\multicolumn{1}{|l|}{Mean CI$^*$ Length} &
  0.88 &
  \multicolumn{1}{c|}{116.77} &
  0.81 &
  \multicolumn{1}{c|}{4882.73} &
  0.66 &
  894.88 \\
\rowcolor[HTML]{EFEFEF} \multicolumn{1}{|l|}{Empirical Coverage} &
  0.98 &
  \multicolumn{1}{c|}{0.89} &
  0.99 &
  \multicolumn{1}{c|}{0.89} &
  0.96 &
  0.99 \\
\multicolumn{1}{|l|}{\begin{tabular}[c]{@{}l@{}}Correct Sign$^{**}$\end{tabular}} &
  0.77 &
  \multicolumn{1}{c|}{0.21} &
  0.89 &
  \multicolumn{1}{c|}{0.3} &
  -- &
  -- \\

\rowcolor[HTML]{EFEFEF} \multicolumn{1}{|l|}{\textbf{Cutoff:} RMSE } &
  0.01 &
  \multicolumn{1}{c|}{--} &
  0.01 &
  \multicolumn{1}{c|}{--} &
  0.05 &
  -- \\
\multicolumn{1}{|l|}{\textbf{Compliance:} RMSE } &

  0.07 &
  \multicolumn{1}{c|}{0.21} &
  0.07 &
  \multicolumn{1}{c|}{0.2} &
  0.08 &
  0.2 \\ \hline
\end{tabular} }
\caption{Simulation results for the fuzzy design. `Known cutoff' means LLR received the true cutoff location as input, while `unknown cutoff' means LoTTA did not have any information about the cutoff. Consequently, the LLR results should be seen as an oracle benchmark and not as a competing method in this setting. $^{*}$`CI' stands for $95\%$ credible interval or confidence interval depending on the context, $^{**}$`Correct sign' refers to the proportion of intervals identifying the sign of the treatment effect.} \label{Tabres23}
\end{table}

\begin{table}

\resizebox{\textwidth}{!}{
\begin{tabular}{|lcccccc|}
\hline
\rowcolor[HTML]{EFEFEF} 
\multicolumn{7}{|l|}{Scenario 2: Fuzzy design, $j=0.55$} \\ \hline
\multicolumn{1}{|l|}{} &
  \multicolumn{2}{c|}{Scenario 2A } &
  \multicolumn{2}{c|}{Scenario 2B } &
  \multicolumn{2}{c|}{Scenario 2C } \\
  \multicolumn{1}{|l|}{} &
  \multicolumn{2}{c|}{(cubic, $\tau=0.31$)} &
  \multicolumn{2}{c|}{(misspecified, $\tau=-0.36$)} &
  \multicolumn{2}{c|}{(misspecified, $\tau=0$)} \\
\multicolumn{1}{|l|}{} &
  \begin{tabular}[c]{@{}c@{}}LoTTA\\ Joint \end{tabular} &
  \multicolumn{1}{c|}{\begin{tabular}[c]{@{}c@{}}LoTTA\\ Treatment-only \end{tabular}} &
  \begin{tabular}[c]{@{}c@{}}LoTTA\\ Joint \end{tabular} &
  \multicolumn{1}{c|}{\begin{tabular}[c]{@{}c@{}}LoTTA\\ Treatment-only \end{tabular}} &
  \begin{tabular}[c]{@{}c@{}}LoTTA\\ Joint \end{tabular} &
  \begin{tabular}[c]{@{}c@{}}LoTTA\\ Treatment-only \end{tabular} \\
  \cline{2-7} 
\rowcolor[HTML]{EFEFEF} \multicolumn{1}{|l|}{\textbf{Cutoff:} RMSE } &
  0.01 &
  \multicolumn{1}{c|}{0.01} &
  0.01 &
  \multicolumn{1}{c|}{0.01} &
  0.02 &
  0.01 \\
 \multicolumn{1}{|l|}{Mean Bias} &
 -0.00 &
  \multicolumn{1}{c|}{0.00} &
  0.00 &
  \multicolumn{1}{c|}{0.00} &
  0.00 &
  0.00 \\
\rowcolor[HTML]{EFEFEF}
\multicolumn{1}{|l|}{Mean CI$^*$ Length} &
  0.02 &
  \multicolumn{1}{c|}{0.05} &
  0.02 &
  \multicolumn{1}{c|}{0.05} &
  0.05 &
  0.05 \\
   \multicolumn{1}{|l|}{Median CI$^*$ Length} &
  0.01 &
  \multicolumn{1}{c|}{0.04} &
  0.01 &
  \multicolumn{1}{c|}{0.04} &
  0.04 &
  0.04 \\
  \rowcolor[HTML]{EFEFEF}
 \multicolumn{1}{|l|}{Empirical Coverage} &
  0.94 &
  \multicolumn{1}{c|}{0.98} &
  0.96 &
  \multicolumn{1}{c|}{0.98} &
  0.96 &
  0.97 \\ \hline

\rowcolor[HTML]{EFEFEF} 
\multicolumn{7}{|l|}{Scenario 3: Fuzzy design, $j=0.3$} \\ \hline
\multicolumn{1}{|l|}{} &
  \multicolumn{2}{c|}{Scenario 3A } &
  \multicolumn{2}{c|}{Scenario 3B } &
  \multicolumn{2}{c|}{Scenario 3C } \\
  \multicolumn{1}{|l|}{} &
  \multicolumn{2}{c|}{(cubic, $\tau=0.57$)} &
  \multicolumn{2}{c|}{(misspecified, $\tau=-0.67$)} &
  \multicolumn{2}{c|}{(misspecified, $\tau=0$)} \\
\multicolumn{1}{|l|}{} &
  \begin{tabular}[c]{@{}c@{}}LoTTA\\ Joint \end{tabular} &
  \multicolumn{1}{c|}{\begin{tabular}[c]{@{}c@{}}LoTTA\\ Treatment-only \end{tabular}} &
  \begin{tabular}[c]{@{}c@{}}LoTTA\\ Joint \end{tabular} &
  \multicolumn{1}{c|}{\begin{tabular}[c]{@{}c@{}}LoTTA\\ Treatment-only \end{tabular}} &
  \begin{tabular}[c]{@{}c@{}}LoTTA\\ Joint \end{tabular} &
  \begin{tabular}[c]{@{}c@{}}LoTTA\\ Treatment-only \end{tabular} \\ \cline{2-7} 
\rowcolor[HTML]{EFEFEF} \multicolumn{1}{|l|}{\textbf{Cutoff:} RMSE } &
  0.01 &
  \multicolumn{1}{c|}{0.04} &
  0.01 &
  \multicolumn{1}{c|}{0.04} &
  0.05 &
  0.04 \\
  \multicolumn{1}{|l|}{Mean Bias} &
  -0.00 &
  \multicolumn{1}{c|}{0.01} &
  0.00 &
  \multicolumn{1}{c|}{0.01} &
  0.01 &
  0.01 \\
\rowcolor[HTML]{EFEFEF}
\multicolumn{1}{|l|}{Mean CI$^*$ Length} &
  0.07 &
  \multicolumn{1}{c|}{0.11} &
  0.06 &
  \multicolumn{1}{c|}{0.11} &
  0.11 &
  0.11 \\
  \multicolumn{1}{|l|}{Median CI$^*$ Length} &
  0.02 &
  \multicolumn{1}{c|}{0.1} &
  0.01 &
  \multicolumn{1}{c|}{0.09} &
  0.09 &
  0.10 \\
  \rowcolor[HTML]{EFEFEF}
 \multicolumn{1}{|l|}{Empirical Coverage} &
  0.95 &
  \multicolumn{1}{c|}{0.96} &
  0.96 &
  \multicolumn{1}{c|}{0.96} &
  0.91 &
  0.96 \\
 \hline
\end{tabular} }
\caption{Simulation results for the cutoff estimation in fuzzy design. `Joint' means that the cutoff was estimated through joint LoTTA model, while `Treatment-only' means that the cutoff was estimated using only treatment model, without influence of the outcome data. $^{*}$`CI' stands for $95\%$ credible interval.} \label{Tabres4}
\end{table}

\section{Discussion}\label{discussion}

We demonstrated through a simulation study and a data application that LoTTA performs well in both cutoff detection and treatment effect estimation. The main strength of the model is that it does not require manual tuning; in particular, a window in which linear parts are fitted is treated as a model's parameter. In addition to being a novel method for analyzing fuzzy RDD's with unknown cutoffs, LoTTA offers competitive performance in the case of a known cutoff and a sharp design. LoTTA credible intervals have good coverage and are relatively narrow. In the case of a non-zero treatment effect, they tend to exclude $0$ with a higher probability than the LLR confidence intervals. Moreover, even though the default version is based on a cubic function, the underlying idea of first-order Taylor approximation near the cutoff can be translated to other functions as well. An example of an extension to the inverse logit function can be found in Appendix \ref{App1}. 

However, since there is no one-size-fits-all solution, our method also comes with some limitations. First, it relies on a cubic (or some kind of) parametrization. While our modification of a cubic function is more robust than the standard polynomial (see Appendix \ref{AppB}), the model can still result in a poor fit. Thus, it is important to investigate the goodness of fit in an analysis, as proposed in Section \ref{diagnostic}. Second, since the method is a Bayesian procedure, it requires significantly more computational time. Additionally, our rather complex model may require a long burn-in period, potentially encountering convergence issues. We recommend to use the Bayesian model with relatively small or noisy datasets, datasets with binary outcomes, and naturally the ones with an unknown or a suspected cutoff. Conversely, in the case of large datasets with continuous outcomes and a known cutoff, local linear regression is a fast and reliable solution. 
Third, while we demonstrated the benefit of the joint model over the cut posterior approach for many settings, we should mention that the latter is likely more robust against violation of Assumption \ref{as3}. While we deem this assumption to be rather realistic, we do advice to visually check it for the data at hand.

One may also wonder whether the point estimate of the cutoff can be directly plugged into an existing RDD estimator, effectively treating the cutoff as known. In Appendix \ref{Plugin} we present a simulation study that shows lack of reliability of such an approach both in terms of point estimates and uncertainty quantification.
\subsection*{Seemingly known cutoff}
The Bayesian model is not only a tool to analyze RD designs with unknown cutoff: it changes the way we view RD data by treating the cutoff as a model parameter. In particular, it challenges the custom of treating cutoffs given by guidelines as known. It is crucial to acknowledge the difference between a cutoff given by a guideline and an empirical cutoff, so the one at which the jump in the treatment probability occurs. Otherwise, a validation procedure may lead to incorrect conclusions and subsequent misleading results or no results at all if the design gets dismissed. Rounding of the scores or mistakes in the communication and/or documentation may cause the cutoff to appear at a different value than expected by researchers. In the case of policy implementation, there might be a delay or haste in the enforcement of a new rule. Possibly, these situations could lead to an invalid RDD.  However, after careful considerations they might as well lead to a valid RDD with a shifted cutoff. LoTTA used as a validation step, may help to detect a shifted cutoff by providing additional information outside the scope of the standard methods. In particular, the resulting posterior distributions give unique insight into the data and plausibility of the design through variety of plots,  allowing for better informed decisions.

\subsection*{Replication code and R package}
Replication code for the results in the main manuscript and in the supplementary materials are available in \texttt{R}, and can be found at \url{https://github.com/JuliaMKowalska/RDD_unknown_cutoff}. An \texttt{R} package implementing our methods is available at:
\url{https://github.com/JuliaMKowalska/LoTTA}

\section{Acknowledgments}
This work is part of the project funded by the Dutch Research Council (NWO) under the Open Competition Domain Science-M programme (agreement No.  OCENW.M20.190).

\newpage
\bibliographystyle{unsrt}  
\bibliography{references}  
\newpage

\appendix
\section{Extensions of the outcome model to bounded and binary outcomes} \label{App1}

 In many applications outcomes take values between $0$ and $1$, and consequently the treatment effect for compliers takes values between $-1$ and $1$. In general, if outcome function is bounded $a\leq f(x) \leq b$ for $x\in\mathcal{I}$, then the treatment effect  $\tau$  takes values in the interval $[a-b,b-a]$ and therefore $j\geq\frac{|\fr{0}-\fl{0}|}{b-a}$, assuming that $j>0$. This knowledge can be included in the joint model through a suitable hierarchical  priors on $j$, $\fl{0}$ and $\fr{0}$: 
\begin{align*}
    &\fl{0}\sim \text{Uniform}(a,b),\\
    &\fr{0}\sim \text{Uniform}(a,b),\\
    &j\sim \text{Uniform}\big(\max\big\{\eta,|\fr{0}-\fl{0}|(b-a)^{-1}\big\},1\big).
\end{align*}

If the outcomes take binary values, the model has to be modified, so that all values of $f$ are between $0$ and $1$. We ensure it through inverse logit link function in the tail, and bounded linear function near the cutoff.

\begin{align}\label{fun_out_bin}
     f_{binary}(x)=
     \begin{cases}
        {\fl{0}}+{\fl{1}}(x-c), & \text{ for } \fk{l}<x<c.\\
        \left(1+e^{-({\fl{\Tilde{0}}}+{\fl{\Tilde{1}}}(x-c)+\fl{2}(x-c)^2+\fl{3}(x-c)^3)}\right)^{-1}, & \text{ for } x\leq \fk{l}<c.\\
        {\fr{0}}+{\fr{1}}(x-c), & \text{ for } c\leq x<\fk{r}.\\
        \left(1+e^{-({\fr{\Tilde{0}}}+{\fr{\Tilde{1}}}(x-c)+\fr{2}(x-c)^2+\fr{3}(x-c)^3)}\right)^{-1}, & \text{ for } c<\fk{r}\leq x.\\
        \end{cases}
\end{align}
In this model, the coefficients $\fl{\Tilde{0}}$,$\fl{\Tilde{1}}$,$\fr{\Tilde{0}}$,$\fr{\Tilde{1}}$ are chosen so that ${\fl{0}}+{\fl{1}}(x-c)$ and ${\fr{0}}+{\fr{1}}(x-c)$ are the first order Taylor approximation at $c$ of $(1+\exp[-({\fl{\Tilde{0}}}+{\fl{\Tilde{1}}}(x-c)+\fl{2}(x-c)^2+\fl{3}(x-c)^3)])^{-1}$ and $(1+\exp[-({\fr{\Tilde{0}}}+{\fr{\Tilde{1}}}(x-c)+\fr{2}(x-c)^2+\fr{3}(x-c)^3)])^{-1}$, respectively. In particular, they are given by
\begin{equation*}
\begin{aligned}[c]
&\fl{\Tilde{0}}=\log\frac{\fl{0}}{1-\fl{0}},\\
&\fl{\Tilde{1}}=\frac{\fl{1}}{\fl{0}(1-\fl{0})},
\end{aligned}
\quad
\begin{aligned}[c]
&\fr{\Tilde{0}}=\log\frac{\fr{0}}{1-\fr{0}},\\
&\fr{\Tilde{1}}=\frac{\fr{1}}{\fr{0}(1-\fr{0})}.
\end{aligned}
\\
\end{equation*}
The priors stay the same as for model \ref{fun_out}, except for the linear coefficients.
\begin{equation*}
\begin{aligned}[c]
&\fl{0}\sim\text{Uniform}(0,1),\\
&\fl{1}\sim\text{Uniform}\Big(\frac{1-\epsilon-\fl{1}}{\fk{l}-c},\frac{\epsilon-\fl{0}}{\fk{l}-c}\Big),
\end{aligned}
\quad
\begin{aligned}[c]
&\fr{0}\sim\text{Uniform}(0,1),\\
&\fr{1}\sim\text{Uniform}\Big(\frac{\epsilon-\fr{0}}{\fk{r}-c},\frac{1-\epsilon-\fr{1}}{\fk{r}-c)^{-1}}\Big).
\end{aligned}
\\
\end{equation*}
\section{Additional Results}\label{AppB}
\subsection{Lee \& Ludwig functions}\label{Lee}

\begin{figure}[h]
    \centering
    \includegraphics[width=\textwidth]{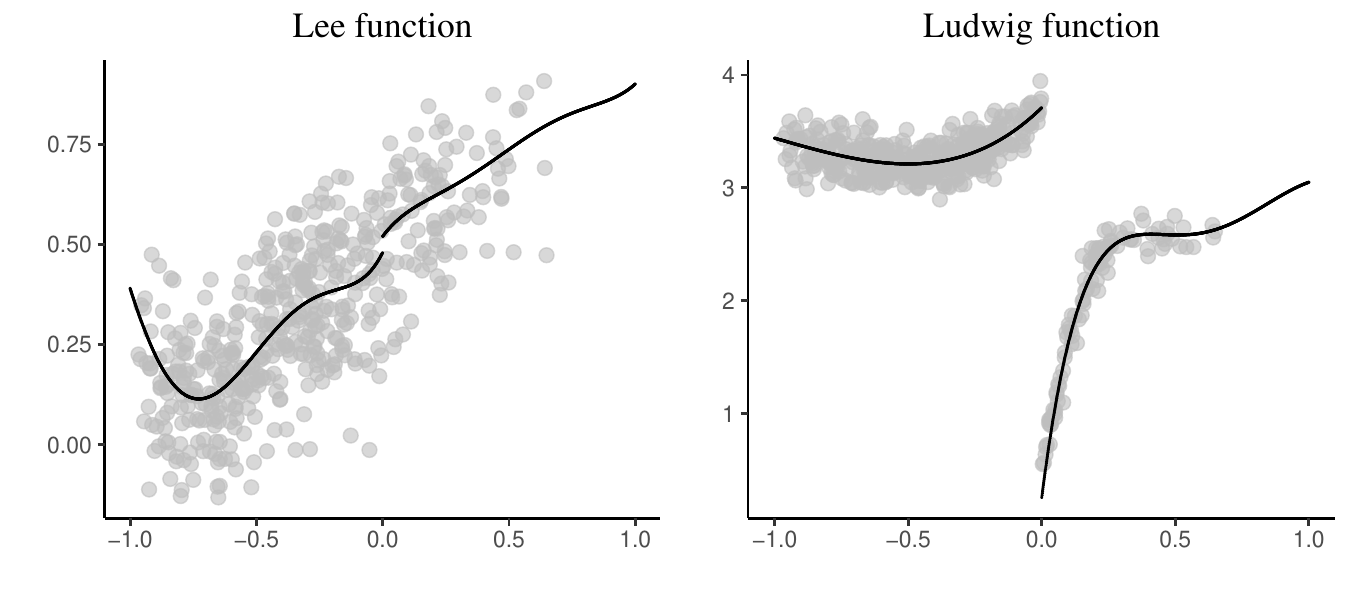}
    \caption{Outcome functions used in simulations: Lee and Ludwig.}
    \label{fig:leelud}
\end{figure}

The functions that are commonly used to compare methods in RDD are Lee and Ludwig functions.
\begin{itemize}
    \item Lee function: $\mu_{Lee}(x)=\mathbbm{1}(x<c)[0.48+ 1.27x+ 7.18x^2 +20.21x^3 + 21.54x^4 + 7.33x^5]+\mathbbm{1}(x\geq c)[0.52+ 0.84x - 3x^2+7.99x^3 -9.01x^4 + 3.56x^5]$.
    \item Ludwig function: $\mu_{Lud}(x)=\mathbbm{1}(x<c)[3.71+ 2.3x+ 3.28x^2 +1.45x^3 + 0.23x^4 + 0.03x^5]+\mathbbm{1}(x\geq c)[0.26+ 18.49x - 54.81x^2+74.3x^3 -45.02x^4 + 9.83x^5]$.
\end{itemize}
Due to their shapes, they are challenging cases (see Figure \ref{fig:leelud}).
In this section we compare LLR, LoTTA and Bayesian cubic regression on 1000 simulated datasets. Cubic regression refers to fitting independently two cubic polynomials on each side of the cutoff. For these  simulations, we set error standard deviation to $0.1295$, so our results can be compared to other results in the RDD literature. \cite{Branson2019}$^,$\cite{Calonico2014}  In Table \ref{leetable}, we see that LoTTA is more robust than the classic cubic model. The biggest pitfall of the cubic model is the low coverage, which is significantly below the nominal value. On the other hand, in the case of the Lee function, LoTTA and LLR both perform similarly well. Both models exhibit some undercoverage; however, LoTTA produces narrower intervals. For the Ludwig function, LLR outperforms the other two models, achieving the best coverage, close to the nominal level, and the smallest bias.  This demonstrates that while LoTTA is more flexible than a simple cubic function, its estimates can still be affected by the model misspecification.
\begin{table}
\centering
\begin{tabular}{|lcccccc|}
\hline
\rowcolor[HTML]{EFEFEF} 
\multicolumn{7}{|l|}{\cellcolor[HTML]{EFEFEF}Sharp design} \\ \hline
\multicolumn{1}{|l|}{} &
  \multicolumn{3}{c|}{Lee ($\tau=0.04$)} &
  \multicolumn{3}{c|}{Ludwig ($\tau=-3.45$)} \\
\multicolumn{1}{|l|}{} &
  \begin{tabular}[c]{@{}c@{}}LoTTA\\ Known \\ Cutoff \end{tabular} &
  \begin{tabular}[c]{@{}c@{}}LLR\\ Known \\ Cutoff \end{tabular} &
  \multicolumn{1}{c|}{\begin{tabular}[c]{@{}c@{}}Cubic\\ Known \\ Cutoff \end{tabular}} &
  \begin{tabular}[c]{@{}c@{}}LoTTA\\ Known \\ Cutoff\end{tabular} &
  \begin{tabular}[c]{@{}c@{}}LLR\\ Known \\ Cutoff\end{tabular} &
  \begin{tabular}[c]{@{}c@{}}Cubic\\ Known \\ Cutoff \end{tabular} \\ \cline{2-7} 
\rowcolor[HTML]{EFEFEF} 
\multicolumn{1}{|l|}{\cellcolor[HTML]{EFEFEF}\textbf{LATE:} RMSE } &
  0.05 &
  0.07 &
  \multicolumn{1}{c|}{\cellcolor[HTML]{EFEFEF}0.1} &
  0.08 &
  0.09 &
  0.1 \\
\multicolumn{1}{|l|}{Average CI Length} &
  0.17 &
  0.25 &
  \multicolumn{1}{c|}{0.19} &
  0.22 &
  0.34 &
  0.2 \\
\rowcolor[HTML]{EFEFEF} 
\multicolumn{1}{|l|}{\cellcolor[HTML]{EFEFEF}Empirical Coverage} &
  0.90 &
  0.91 &
  \multicolumn{1}{c|}{\cellcolor[HTML]{EFEFEF}0.59} &
  0.84 &
  0.93 &
  0.63 \\
\multicolumn{1}{|l|}{\begin{tabular}[c]{@{}l@{}}Proportion Correct Sign$^*$\end{tabular}} &
  0.32 &
  0.16 &
  \multicolumn{1}{c|}{0.73} &
  1 &
  1 &
  1 \\
\rowcolor[HTML]{EFEFEF} 
\multicolumn{1}{|l|}{\cellcolor[HTML]{EFEFEF}Mean Bias} &
  0.02 &
  0.01 &
  \multicolumn{1}{c|}{\cellcolor[HTML]{EFEFEF}0.09} &
  0.04 &
  0.00 &
  0.08 \\ \hline
\end{tabular}%

\caption{Comparison of LoTTA, LLR, and cubic regression for Lee and Ludwig functions. $^*$ `Correct sign' refers to intervals identifying the sign of the treatment effect. }
\label{leetable}
\end{table}
\subsection{Plugin estimator with naive confidence intervals}\label{Plugin}
In this section, we use a simulation study to illustrate the consequences of directly plugging the point estimate of the cutoff into an LLR estimator without accounting for the additional uncertainty. The functions we used for this simulation study are the same as in Section \ref{simulations}. We simulated 1000 datasets of 500 datapoints. In each dataset we estimate the cutoff by taking the MAP estimate from both two constant functions and the LoTTA treatment model. Then we plug it in the local linear regression estimator (LLR) from the \texttt{rdrobust} package to  estimate the treatment effect and robust $95\%$ confidence intervals. We present the results in Table \ref{Tabres6}. They clearly show that, unless there is no treatment effect, the plugin estimator leads to increased bias and unreliable uncertainty quantification (cf. Table \ref{Tabres23})
\begin{table}

\resizebox{\textwidth}{!}{
\begin{tabular}{|lcccccc|}
\hline
\rowcolor[HTML]{EFEFEF} 
\multicolumn{7}{|l|}{Scenario 2: Fuzzy design, $j=0.55$} \\ \hline
\multicolumn{1}{|l|}{} &
  \multicolumn{2}{c|}{Scenario 2A } &
  \multicolumn{2}{c|}{Scenario 2B } &
  \multicolumn{2}{c|}{Scenario 2C } \\
  \multicolumn{1}{|l|}{} &
  \multicolumn{2}{c|}{(cubic, $\tau=0.31$)} &
  \multicolumn{2}{c|}{(misspecified, $\tau=-0.36$)} &
  \multicolumn{2}{c|}{(misspecified, $\tau=0$)} \\
\multicolumn{1}{|l|}{} &
  \begin{tabular}[c]{@{}c@{}} Two\\ Constant \end{tabular} &
  \multicolumn{1}{c|}{\begin{tabular}[c]{@{}c@{}}LoTTA\\ Treatment-only \end{tabular}} &
  \begin{tabular}[c]{@{}c@{}}Two\\ Constant \end{tabular} &
  \multicolumn{1}{c|}{\begin{tabular}[c]{@{}c@{}}LoTTA\\ Treatment-only \end{tabular}} &
  \begin{tabular}[c]{@{}c@{}}Two\\ Constant \end{tabular} &
  \begin{tabular}[c]{@{}c@{}}LoTTA\\ Treatment-only \end{tabular} \\
  \cline{2-7} 
\rowcolor[HTML]{EFEFEF} \multicolumn{1}{|l|}{\textbf{LATE:} RMSE } &
  0.18 &
  \multicolumn{1}{c|}{0.15} &
  0.18 &
  \multicolumn{1}{c|}{0.17} &
  0.08 &
  0.08 \\
 \multicolumn{1}{|l|}{Mean Bias} &
  -0.07 &
  \multicolumn{1}{c|}{-0.07} &
  0.08 &
  \multicolumn{1}{c|}{0.08} &
  0.00 &
  0.00 \\
\rowcolor[HTML]{EFEFEF}
\multicolumn{1}{|l|}{Mean CI Length} &
  0.51 &
  \multicolumn{1}{c|}{0.48} &
  0.50 &
  \multicolumn{1}{c|}{0.49} &
  0.32 &
  0.32 \\
 \multicolumn{1}{|l|}{Empirical Coverage} &
  0.79 &
  \multicolumn{1}{c|}{0.79} &
  0.75 &
  \multicolumn{1}{c|}{0.74} &
  0.96 &
  0.96 \\ \hline

\rowcolor[HTML]{EFEFEF} 
\multicolumn{7}{|l|}{Scenario 3: Fuzzy design, $j=0.3$} \\ \hline
\multicolumn{1}{|l|}{} &
  \multicolumn{2}{c|}{Scenario 3A } &
  \multicolumn{2}{c|}{Scenario 3B } &
  \multicolumn{2}{c|}{Scenario 3C } \\
  \multicolumn{1}{|l|}{} &
  \multicolumn{2}{c|}{(cubic, $\tau=0.57$)} &
  \multicolumn{2}{c|}{(misspecified, $\tau=-0.67$)} &
  \multicolumn{2}{c|}{(misspecified, $\tau=0$)} \\
\multicolumn{1}{|l|}{} &
  \begin{tabular}[c]{@{}c@{}}Two\\ Constant \end{tabular} &
  \multicolumn{1}{c|}{\begin{tabular}[c]{@{}c@{}}LoTTA\\ Treatment-only \end{tabular}} &
  \begin{tabular}[c]{@{}c@{}}Two\\ Constant \end{tabular} &
  \multicolumn{1}{c|}{\begin{tabular}[c]{@{}c@{}}LoTTA\\ Treatment-only \end{tabular}} &
  \begin{tabular}[c]{@{}c@{}}Two\\ Constant \end{tabular} &
  \begin{tabular}[c]{@{}c@{}}LoTTA\\ Treatment-only \end{tabular} \\ \cline{2-7} 
\rowcolor[HTML]{EFEFEF} \multicolumn{1}{|l|}{\textbf{LATE:} RMSE } &
  0.42 &
  \multicolumn{1}{c|}{0.41} &
  0.49 &
  \multicolumn{1}{c|}{0.48} &
  0.14 &
  0.12 \\
  \multicolumn{1}{|l|}{Mean Bias} &
  -0.33 &
  \multicolumn{1}{c|}{-0.32} &
  0.40 &
  \multicolumn{1}{c|}{0.40} &
  0.01 &
  0.01 \\
\rowcolor[HTML]{EFEFEF}
\multicolumn{1}{|l|}{Mean CI Length} &
  0.87 &
  \multicolumn{1}{c|}{0.85} &
  0.89 &
  \multicolumn{1}{c|}{0.83} &
  0.55 &
  0.50 \\
 \multicolumn{1}{|l|}{Empirical Coverage} &
  0.50 &
  \multicolumn{1}{c|}{0.49} &
  0.42 &
  \multicolumn{1}{c|}{0.42} &
  0.98 &
  0.99 \\
 \hline
\end{tabular} }
\caption{Simulation results for the plugin estimator. `Two constant' means that the cutoff was estimated by fitting two constant functions and taking MAP of the posterior distribution. `LoTTA Treatment-only' means that the cutoff was estimated with LoTTA treatment model. } \label{Tabres6}
\end{table}

\subsection{Chemotherapy dataset}\label{chemo}
\begin{figure}
    \centering
    \includegraphics[width=\textwidth]{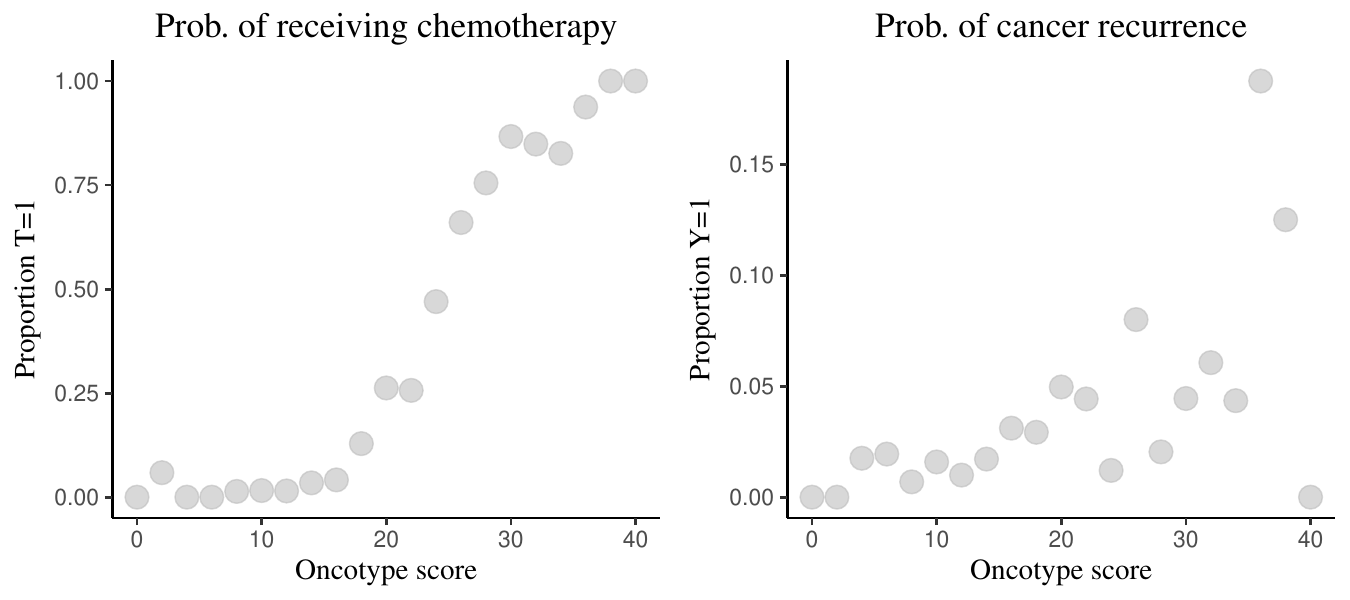}
    \caption{Binned data of chemotherapy take-up and cancer recurrence. There is visible increase in the chemotherapy take-up in the interval $[20,30]$. }
    \label{fig:chemo_data}
\end{figure}

\begin{figure}
    \centering
    \includegraphics[width=\textwidth]{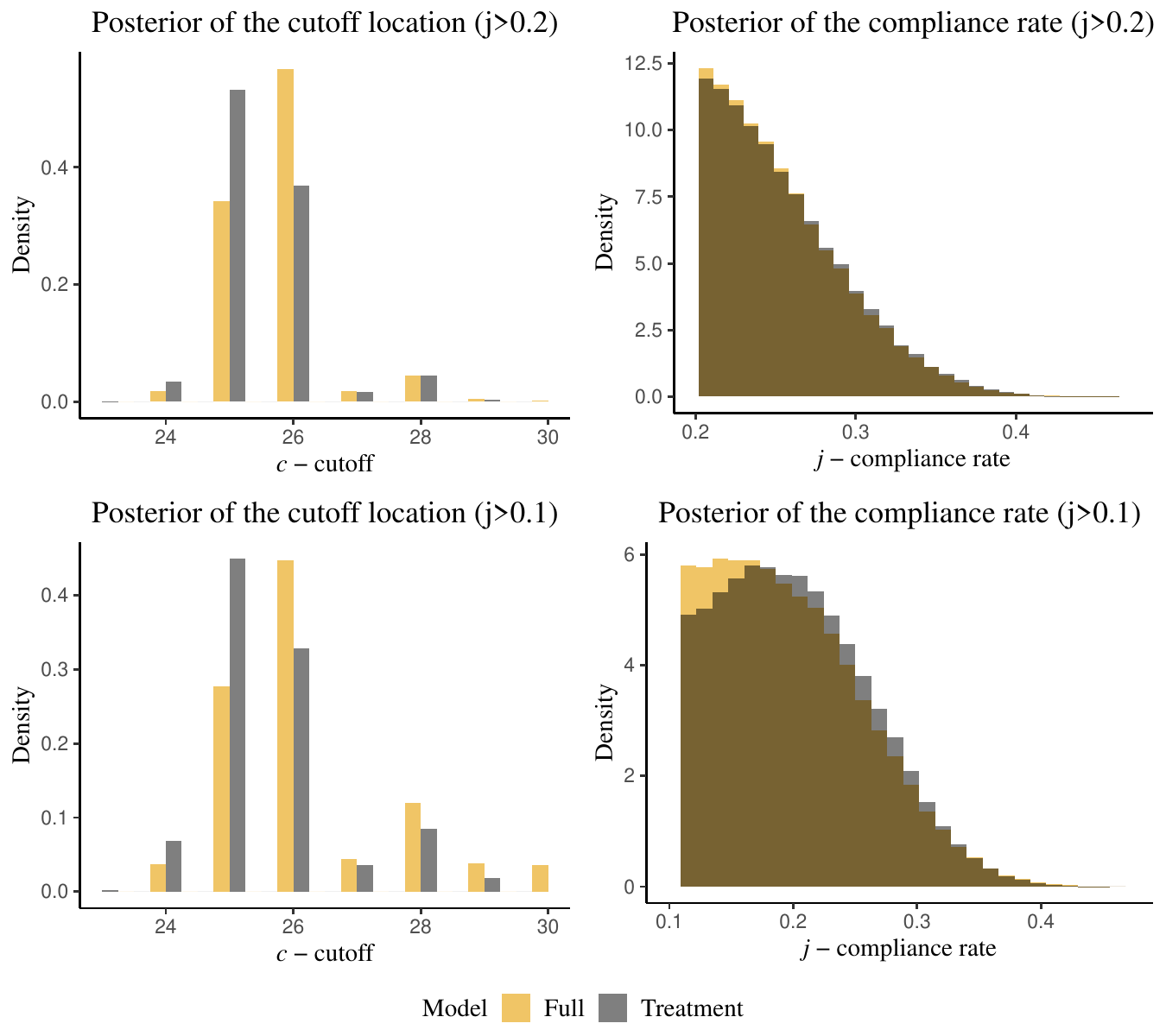}
    \caption{Posterior distributions of the cutoff location and compliance rate of the chemotherapy take-up. The posterior of the full model puts the highest mass on the suspected cutoff of $26$, but the neighboring point also have high posterior probability of being a cutoff point. Together with posterior distribution skewed to the left they suggest invalid RDD. }
    \label{fig:chemo_out}
\end{figure}

Cattaneo et al. in their paper \textit{A Guide to Regression Discontinuity Designs in Medical Applications}\cite{Cattaneo2023} give an example of a faulty regression discontinuity design. The data the authors provide,\cite{Chemo2023} contains information about oncoscore of patients, chemotherapy uptake, and cancer re-occurence. A common guideline  is to give chemotherapy to patients with oncoscore equal or higher than 26. This creates a potential opportunity to estimate the effect of the chemotherapy through a fuzzy RDD. However, the authors' analysis shows that the compliance rate is too low and crucial covariates are not balanced near the cutoff. Therefore, the  design is not valid. We can come to the same conclusion through the Bayesian analysis as well. We start by plotting the treatment allocation data, see Figure \ref{fig:chemo_data}. We observe that the treatment take-up increases in the interval between the scores $20$ and $30$. We set uniform prior on the cutoff location between $20$ and $30$,  and we fit the treatment and the full model with lower bound of $\eta=0.2$ on the compliance rate. The histogram shows that posterior concentrates at two neighbouring points $26$ and $25$ and jump in the treatment probability function leans towards the lower bound, see Figure \ref{fig:chemo_out}. Such posterior distributions of $c$ and $j$ suggest a steep increase instead of a jump. Next, we lower the bound on the compliance rate to $0.1$. Again, the posterior concentrates around the same points, but this time the posterior of the compliance rate has a visible peak around $0.19$. Nonetheless, there is still considerable mass near $0.1$. Even though, for both choices of $\eta$, the full model put the highest mass on the suspected cutoff of $26$, the neighboring point also have significant probability of being a cutoff point. We conclude that the design is not valid.

\subsection{ART dataset: sensitivity analysis}\label{APP:ART_sens}
In this section we present the results the detailed results of the sensitivity analysis on ART dataset. We investigated sensitivity of our results to $\eta$, which is the lower bound on the compliance rate, and prior on the cutoff location.

\begin{figure}
    \centering
    \includegraphics[width=\textwidth]{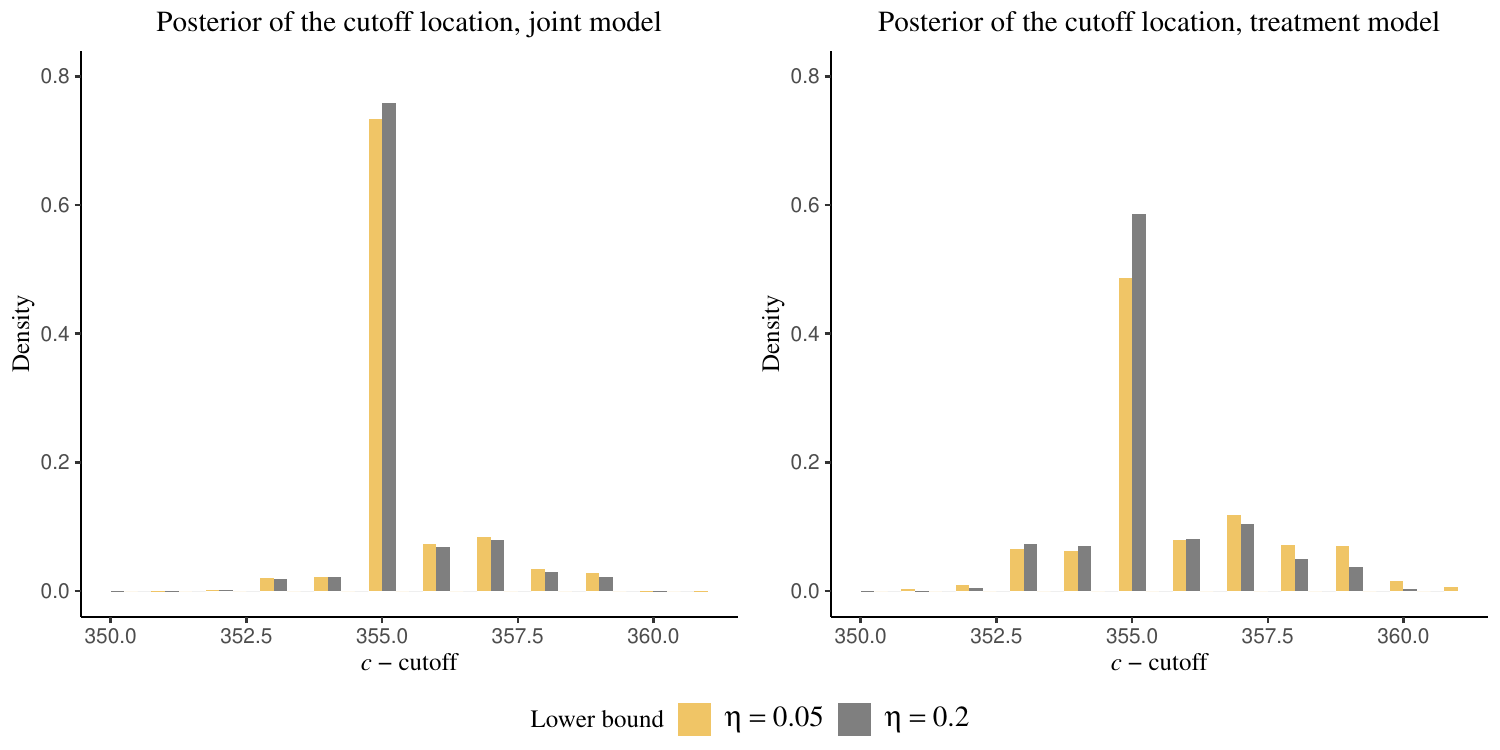}\\
    \includegraphics[width=\textwidth]
    {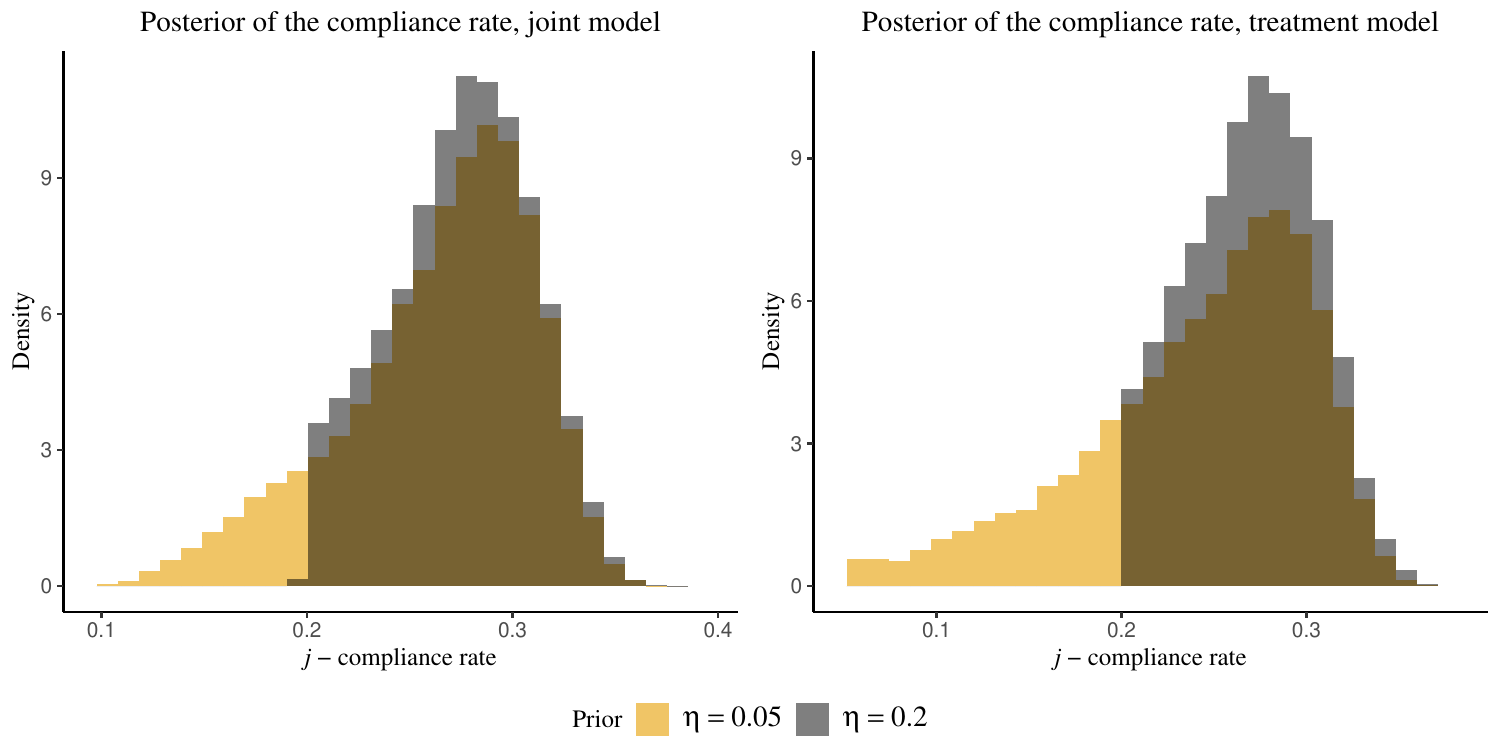}
    \caption{Comparison of posterior distributions for cutoff location and compliance rate under $\eta = 0.05$ and $\eta = 0.2$ in the ART dataset. }
    \label{fig:artsens_eta}
\end{figure}

To examine the influence of $\eta$, we refitted both the treatment and full LoTTA models using $\eta = 0.05$ and compared the results to those obtained with $\eta = 0.2$ in the original analysis. As shown in Figure \ref{fig:artsens_eta}, the posterior distribution of the cutoff location remains concentrated around $355$ in both cases. As expected, the posterior distribution of the compliance rate differs between the two $\eta$ values, particularly in the left tail. However, as reported in Table \ref{table:artsens}, the 95\% credible intervals remain well separated from 0.05 for both the joint and treatment models. This provides supporting evidence for the validity of the design. The 95\% credible interval for the treatment effect widens under $\eta = 0.05$. Specifically, allowing for lower compliance rates shifts  posterior mass toward higher treatment effects. At the same time the point estimates of LATE are almost equal for both $\eta$'s.

\begin{figure}
    \centering
    \includegraphics[width=\textwidth]{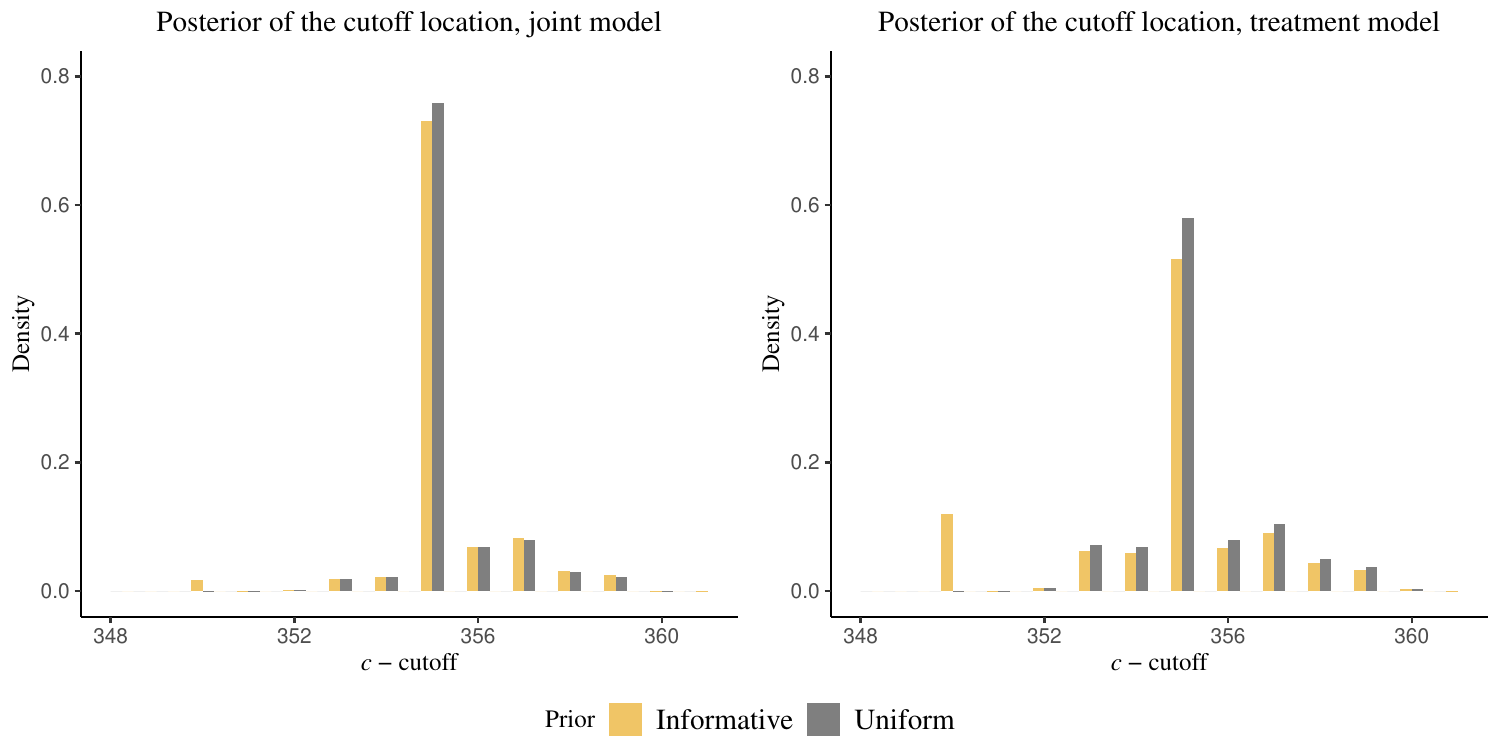}\\
    \includegraphics[width=\textwidth]
    {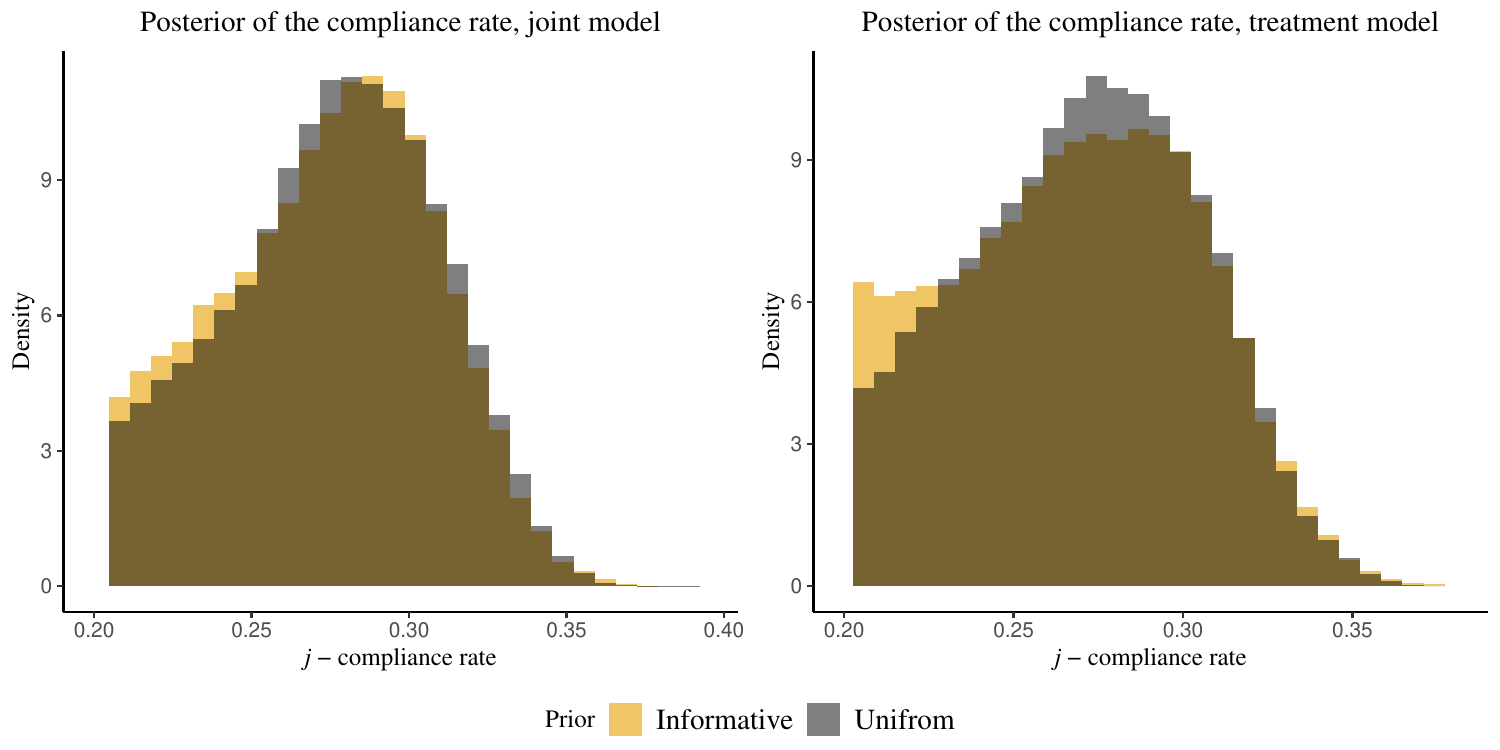}
    \caption{Comparison of posterior distributions for cutoff location and compliance rate under uniform and strongly informative prior on the cutoff location in the ART dataset. }
    \label{fig:artsens_prior}
\end{figure}

To examine the influence of the prior on the cutoff location, we refitted both the treatment and full LoTTA models with strongly informative prior. Specifically, we set a prior that assigns $90\%$ of the prior mass to $350$, the official cutoff. The remaining $10\%$ is distributed uniformly over all other cutoff values in the interval $[300, 399]$.  In the original analysis we assumed uniform prior on that interval. As we see on Figure \ref{fig:artsens_prior}, in the treatment-only model there is still substantial posterior mass on $350$, however most of the posterior mass is on $355$. In the joint model, the informative and uniform priors lead to very similar posterior distributions for both the cutoff location and compliance rate. Consequently, the estimates of LATE are also nearly identical in terms of the point estimates and credible intervals (Table \ref{table:artsens}).

\begin{table}[ht]
\centering
\resizebox{\textwidth}{!}{
\begin{tabular}{|lccc|}

\rowcolor[HTML]{EFEFEF} 
\hline
\multicolumn{4}{|l|}{Lower bound $\eta$}                                                                                                              \\ \hline
\multicolumn{1}{|l|}{}                               & \multicolumn{2}{c|}{$\eta=0.05$}                                                            & $\eta=0.2$                 \\ 
\multicolumn{1}{|l|}{}                               & \multicolumn{1}{c}{Treatment model}               & \multicolumn{1}{c|}{Joint model}               & \multicolumn{1}{l|}{Joint model} \\ \cline{2-4} 
\multicolumn{1}{|l|}{LATE} & \multicolumn{1}{c}{--} & \multicolumn{1}{c|}{-0.55(-0.94,-0.38)} & -0.56 (-0.85,-0.38)   \\ 
\multicolumn{1}{|l|}{Compliance rate}                & \multicolumn{1}{c}{0.28 (0.11,0.34)}    & \multicolumn{1}{c|}{0.29 (0.16,0.34)}    & 0.28 (0.2,0.33)       \\ 
\multicolumn{1}{|l|}{Cutoff location}                & \multicolumn{1}{c}{355 [353,360]}                   & \multicolumn{1}{c|}{355 [355,359]}                   & 355 [354,358]         \\ 

\rowcolor[HTML]{EFEFEF} 
\hline
\multicolumn{4}{|l|}{Prior on the cutoff location}                                                                                                          \\ \hline
\multicolumn{1}{|l|}{}                               & \multicolumn{2}{c|}{Strongly informative}                                                            & Uniform                 \\ 
\multicolumn{1}{|l|}{}                               & \multicolumn{1}{c}{Treatment model}               & \multicolumn{1}{c|}{Joint model}               & \multicolumn{1}{l|}{Joint model} \\ \cline{2-4} 
\multicolumn{1}{|l|}{LATE} & \multicolumn{1}{c|}{--}  & \multicolumn{1}{c|}{-0.56 (-0.83,-0.37)} & -0.56 (-0.85,-0.38)   \\  
\multicolumn{1}{|l|}{Compliance rate}                & \multicolumn{1}{c|}{0.28 (0.2,0.32)}   & \multicolumn{1}{c|}{0.29 (0.2,0.33)}    & 0.28 (0.2,0.33)       \\ 
\multicolumn{1}{|l|}{Cutoff location}                & \multicolumn{1}{c|}{355 [350,358]}                   & \multicolumn{1}{c|}{355 [354,359]}                   & 355 [354,358]         \\ \hline
\end{tabular}
}
\caption{ Sensitivity analysis of LoTTA parameters on ART data. ART dataset was initially analysed with $\eta=0.2$ and uniform prior on the cutoff location between 300 and 399. For each estimand, MAP estimates are given along with 95\% highest density intervals.}
\label{table:artsens}
\end{table}
\section{Cut posterior}\label{cut}
In this section, we compare the joint posterior with the cut posterior on two simulated datasets of $500$ data points with and without treatment effect. The two datasets share the same treatment data, but have different outcome data. The treatment data was generated according to the function $p_2$ given in Section \ref{simulations} that corresponds to the compliance rate of $0.3$. In the first dataset outcomes were generated according to the function $\mu_B$ with the discontinuity at $0$ of size $-0.2$. In the second dataset, outcomes were generated according to the continuous function $\mu_C$. 

Since the compliance rate is only 0.3, the treatment data does not contain a strong signal for the cutoff location, and some datasets suggest that the cutoff is at 0.05 instead of the true value of 0.

\begin{figure}
    \centering
    \includegraphics[width=\textwidth]{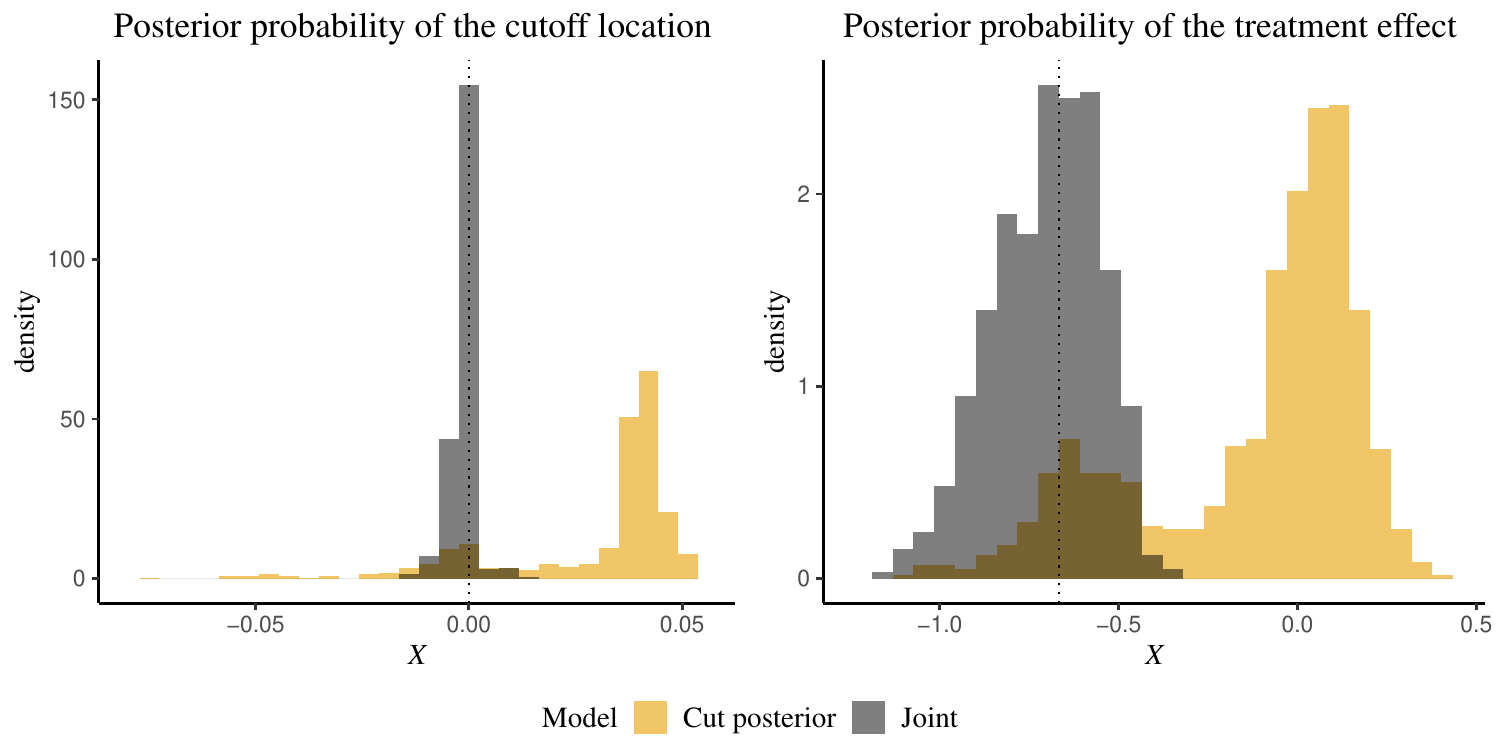}
    \caption{Histogram of 1000 posterior samples of the cutoff location and treatment effect from a single dataset with negative treatment effect. The true values are marked with dotted lines. Low compliance rate of $0.3$ and little data lead to misleading results regarding the cutoff location if only treatment allocation data is considered. Consequently, in the cut posterior approach, the error is propagated to the treatment effect estimation.       }
    \label{fig:cutposterior02}
\end{figure}

Function $\mu_B$ does contain information about the cutoff location, and we observe that the joint model correctly detects the cutoff at 0 and its posterior concentrates around the true treatment effect (Figure \ref{fig:cutposterior02}). The cut posterior cannot use the signal in $\mu_B$ and incorrectly concentrates on a cutoff location of 0.05, which then leads to a posterior on the treatment effect that is mostly located away from the true value.

\begin{figure}
    \centering
    \includegraphics[width=\textwidth]{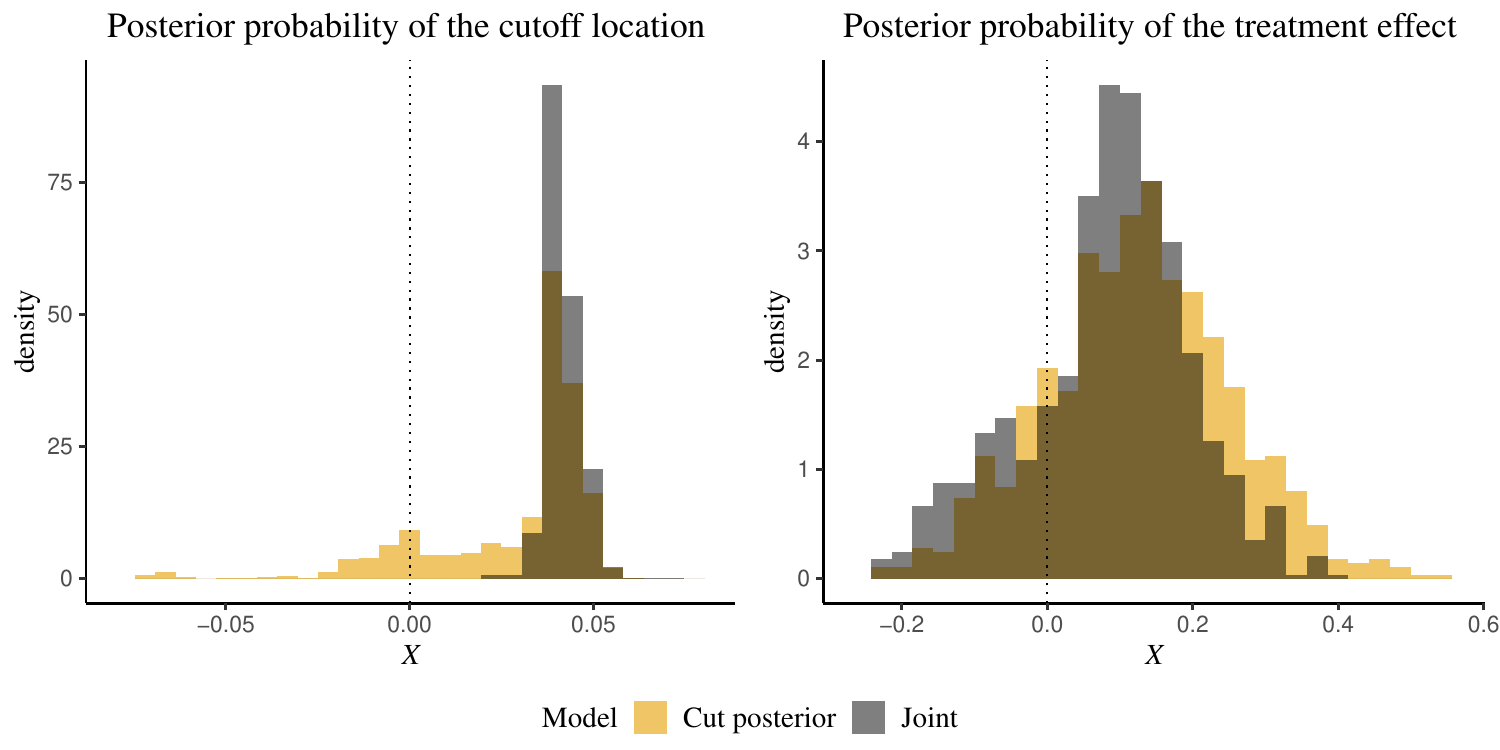}
    \caption{Histogram of 1000 posterior samples of the cutoff location and treatment effect from a single dataset with no treatment effect. The true values are marked with dotted lines. In case of no treatment effect the joint model does not have additional information to correct the cutoff location. }
    \label{fig:cutposterior00}
\end{figure}

Function $\mu_C$ contains no information about the cutoff location, and here we observe similar behavior of the joint model and cut posterior (Figure \ref{fig:cutposterior00}). Both posteriors concentrate around an incorrect cutoff location of 0.05. In this case both posteriors on the treatment effect do center around the correct value of 0, since $\mu_C$ is continuous everywhere and thus even at a wrong cutoff location, a correct treatment effect of 0 will be found.

In summary, the cut posterior only uses the signal in the treatment data to find the cutoff location, while the joint posterior can use the treatment data as well as the outcome data to find the correct location, and especially in Figure \ref{fig:cutposterior02} we see that the opportunity to use this additional source of information leads to better results.
\section{Identification result for multiple cutoffs}\label{App:iden_res}
In the case of an unknown or a shifted cutoff, there is a concern that the detected cutoff $c$ was not used consistently by all decision makers. This raises the question of whether the treatment effect remains identifiable and under which conditions. In this section, we demonstrate that, under mild assumptions analogous to those in the standard fuzzy RDD framework, the treatment effect is still identifiable. However, the interpretation changes: in this case the treatment effect is estimated for compliers with scores equal to $c$, whose treatment was decided through the cutoff rule $c$. 

First, we introduce a variable $R$ that corresponds to the cutoff value used by a random decision maker. 
\begin{assumption}\label{as1_app}
    $R$ takes value in a finite set $\{c_1,\ldots,c_n\}$, where $c_i<c_{i+1}$.
\end{assumption}
Consequently, we assume that each decision maker used some cutoff value to determine the treatment. Our results generalize to the case in which fraction of decision makers did not use any cutoff rule but for the clarity of exposure we omit this scenario. 

As previously, we assume that there are no defiers.
\begin{assumption}\label{as2_app}
    $\mathbb{P}[T=1|X=x_1,R=c_i]\leq \mathbb{P}[T=1|X=x_2,R=c_i]$ for $x_1<x_2$ and $i=1,\ldots,n$.
\end{assumption}

Second, we consider potential outcomes of the form $Y^{(t,r)}$, which represent the outcome a patient would have if they received treatment $t$ and were treated by a decision maker who used cutoff rule $r$. This formulation allows for heterogeneity across decision makers based on the cutoff rule they employed. The underlying premise is that decision makers who deviated from the official cutoff may differ systematically from those who adhered to it.  For example, in medical studies, doctors who used a shifted cutoff might have paid more attention to their patients, or if the data was collected at multiple locations, they might have been working at a particular clinic. 

Now, we need to establish the way we think of compliance types. We consider compliance type to be a baseline characteristic, that depends on both observed and unobserved variables. In our interpretation the always and never-takers are the units for which the treatment is decided upfront due to available knowledge or strong preferences of those units. Formally, $\mathbb{P}(T=1|C_T=A,X=x,R=r)=1$ and $\mathbb{P}(T=1|C_T=N,X=x,R=r)=0$. The compliers on the other hand, are individuals for whom the available information and current state of knowledge is not sufficient to determine treatment directly. For them, the treatment decision follows a cutoff rule. That is, $\mathbb{P}(T=1|C_T=C,X=x,R=r)=1$  if $x\geq r$ and $\mathbb{P}(T=1|C_T=C,X=x,R=r)=0$  if  $x< r$. We also assume that each decision maker applies a fixed cutoff rule consistently to all individuals they assess. These ideas are summarized in the following assumption.
\begin{assumption}\label{as3_app}
    $\mathbb{P}(C_T=c_T|X=x,R=r)$ and $\mathbb{P}(R=r|X=x)$ are continuous in $x$ for each compliance type $c_T=A,N,C$ and cutoff rule $r\in\{c_1,\ldots,c_n\}$.
\end{assumption}
We can view this assumption as similar to the standard RDD assumption that individuals cannot precisely manipulate their score to influence treatment. The first part implies that there are no abrupt changes in the compliance process. As an example, it would be violated if individuals systematically manipulated their score in order to receive the treatment they prefer. In the second part we assume that individuals do not select their decision maker in order to manipulate the treatment they receive.

To derive the identification result we need to formulate continuity assumption analogous to the standard RDD set up. 
\begin{assumption}\label{as4_app}
    Each conditional expected value $\mathbb{E}[Y^{(i,r)}|X=x,R=r,C_T=c_t]$ is continuous in $x$ for each cutoff rule $r\in \{c_1,\ldots,c_n\} $ and compliance type $c_T=\text{A}, \text{N}, \text{C}$.
\end{assumption}
Altogether, the above assumptions ensure that any discontinuity in outcomes at a cutoff can be attributed solely to the treatment.
\begin{theorem}
 Let Assumptions \ref{as1_app}--\ref{as4_app} hold. Then the local average treatment effect is identified at each $c_i$ for the subgroup of compliers, whose treatment was determined according to the cutoff $c_i$.
 \begin{align}\label{res1_app}
     \mathbb{E}[Y^{(1,c_i)}-Y^{(0,c_i)}|X=c_i,C_T=C,R=c_i]=\frac{\lim_{x\downarrow c_i}\mathbb{E}[Y|X=x]-\lim_{x\uparrow c_i}\mathbb{E}[Y|X=x]}{\lim_{x\downarrow c_i}\mathbb{P}[T=1|X=x]-\lim_{x\uparrow c_i}\mathbb{P}[T=1|X=x]}.
 \end{align}
    Moreover, if we assume that $Y^{(t,c_i)}\perp R | X,C$ for $t=0,1$, then 
    \begin{align}\label{res2_app}
        \mathbb{E}[Y^{(1,c_i)}-Y^{(0,c_i)}|X=c_i,C_T=C,R=c_i]=\mathbb{E}[Y^{(1,c_i)}-Y^{(0,c_i)}|X=c_i,C_T=C].
    \end{align}
    Additionally, if we also assume $\mathbb{E}[Y^{(t,r_1)}|X=x,C_t=C]=\mathbb{E}[Y^{(t,r_2)}|X=x,C_t=C]$ for any $r_1$, $r_2\in\{c_1,\ldots,c_n\}$, then
    \begin{align}\label{res3_app}
        \mathbb{E}[Y^{(1,c_i)}-Y^{(0,c_i)}|X=c_i,C_T=C]=\mathbb{E}[Y^{(1)}-Y^{(0)}|X=c_i,C_T=C].
    \end{align}
\end{theorem}
 \begin{proof}
        The proof is analogous to the proof for the standard identification result in fuzzy RDD. 
        
         Let's fix $c_i$ and $x\in(c_{i-1},c_{i})$. By Assumptions \ref{as1_app} and \ref{as2_app}, we decompose $\mathbb{E}[Y|X=x]$ in the following way.
       
        \begin{align*}
          &\mathbb{E}[Y|X=x]=\\
          &=\sum_{c_T\in\{A,N,C\}}\sum_{j=1}^n\mathbb{E}[Y|X=x,R=c_j,C_T=c_T]\mathbb{P}[C_T=c_T|X=x,R=c_j]\mathbb{P}(R=c_j|X=x)\\
          &=\sum_{j=1}^n\mathbb{E}[Y^{(1,c_j)}|X=x,R=c_j,C_T=A]\mathbb{P}[C_T=A|X=x,R=c_j]\mathbb{P}(R=c_j|X=x)  \\
          &+\sum_{j=1}^n\mathbb{E}[Y^{(0,c_j)}|X=x,R=c_j,C_T=N]\mathbb{P}[C_T=N|X=x,R=c_j]\mathbb{P}(R=c_j|X=x)\\
          &+\sum_{j=1}^{i-1}\mathbb{E}[Y^{(1,c_j)}|X=x,R=c_j,C_T=C]\mathbb{P}[C_T=C|X=x,R=c_j]\mathbb{P}(R=c_j|X=x)\\
          &+\sum_{j=i}^{n}\mathbb{E}[Y^{(0,c_j)}|X=x,R=c_j,C_T=C]\mathbb{P}[C_T=C|X=x,R=c_j]\mathbb{P}(R=c_j|X=x).
        \end{align*}
        Then by taking the limit $x\uparrow c_i$ and using Assumption \ref{as3_app} and \ref{as4_app}, we get
        \begin{align*}
          &\lim_{x\uparrow c_i}\mathbb{E}[Y|X=x]=\\
          &=\sum_{j=1}^n\mathbb{E}[Y^{(1,c_j)}|X=c_i,R=c_j,C_T=A]\mathbb{P}[C_T=A|X=c_i,R=c_j]\mathbb{P}(R=c_j|X=c_i) \\
          &+\sum_{j=1}^n\mathbb{E}[Y^{(0,c_j)}|X=c_i,R=c_j,C_T=N]\mathbb{P}[C_T=N|X=c_i,R=c_j]\mathbb{P}(R=c_j|X=c_i)\\
          &+\sum_{j=1}^{\boldsymbol{i-1}}\mathbb{E}[Y^{(1,c_j)}|X=c_i,R=c_j,C_T=C]\mathbb{P}[C_T=C|X=c_i,R=c_j]\mathbb{P}(R=c_j|X=c_i)\\
          &+\sum_{j=\boldsymbol{i}}^{n}\mathbb{E}[Y^{(0,c_j)}|X=c_i,R=c_j,C_T=C]\mathbb{P}[C_T=C|X=c_i,R=c_j]\mathbb{P}(R=c_j|X=c_i).
        \end{align*}
    Similarly, for $x\in[c_i,c_{i+1})$ we obtain
    \begin{align*}
          &\lim_{x\downarrow c_i}\mathbb{E}[Y|X=x]=\\
          &=\sum_{j=1}^n\mathbb{E}[Y^{(1,c_j)}|X=c_i,R=c_j,C_T=A]\mathbb{P}[C_T=A|X=c_i,R=c_j]\mathbb{P}(R=c_j|X=c_i) \\
          &+\sum_{j=1}^n\mathbb{E}[Y^{(0,c_j)}|X=c_i,R=c_j,C_T=N]\mathbb{P}[C_T=N|X=c_i,R=c_j]\mathbb{P}(R=c_j|X=c_i)\\
          &+\sum_{j=1}^{\boldsymbol{i}}\mathbb{E}[Y^{(1,c_j)}|X=c_i,R=c_j,C_T=C]\mathbb{P}[C_T=C|X=c_i,R=c_j]\mathbb{P}(R=c_j|X=c_i)\\
          &+\sum_{j=\boldsymbol{i+1}}^{n}\mathbb{E}[Y^{(0,c_j)}|X=c_i,R=c_j,C_T=C]\mathbb{P}[C_T=C|X=c_i,R=c_j]\mathbb{P}(R=c_j|X=c_i).
        \end{align*} Then by taking the difference and using Bayes' rule we obtain:
        \begin{align*}
    &\lim_{x\downarrow c_i} \mathbb{E}[Y \mid X=x] - \lim_{x\uparrow c_i} \mathbb{E}[Y \mid X=x] \\
    &= \left( \mathbb{E}[Y^{(1,c_i)} \mid X=c_j, R=c_i, C_T=C] - \mathbb{E}[Y^{(0,c_i)} \mid X=c_j, R=c_j, C_T=C] \right) \\
    &\quad \times \mathbb{P}[C_T=C, R=c_i \mid X=c_i]
\end{align*}
Again, in the same spirit we obtain
       \begin{align*}
          &\lim_{x\downarrow c_i}\mathbb{P}(T=1|X=x)=\mathbb{P}(C_T=C,R \leq c_i|X=c_i)+\mathbb{P}(C_T=A|X=c_i)\\
          &\lim_{x\uparrow c_i}\mathbb{P}(T=1|X=x)=\mathbb{P}(C_T=C,R < c_i|X=c_i)+\mathbb{P}(C_T=A|X=c_i)
       \end{align*}
       which gives us the final result by taking the difference. The remaining statements follow directly from the additional assumptions.
    \end{proof}

The denominator in the formula \eqref{res1_app} does not equal the compliance rate at $c_j$ as in a standard RDD. Instead, it reflects the proportion of units at $c_j$ who are compliers and whose treatment was determined according to the rule $R=c_j$. 

Moreover, we would like to stress that we do not claim that it is feasible to estimate the treatment effect at each cutoff value. Indeed, we focus on the scenario in which the values $\lim_{x\downarrow c_i}\mathbb{P}[T=1|X=x]-\lim_{x\uparrow c_i}\mathbb{P}[T=1|X=x]$ are small for all $i$ except for one. For this reason, we do not expect that the alternative cutoff values influence the estimated values in the numerator and denominator in formula \eqref{res1_app}. We work under the assumption that the proportion of the decision makers using each alternative cutoff is relatively small compared to the sample size, hence those cutoffs are not detectable. With binary treatment allocation data being little informative it is a plausible assumption. 

\paragraph{Interpretation of the results.} We are aware that the differences between the quantities \eqref{res1_app}, \eqref{res2_app} and \eqref{res3_app} may seem obscure at first. To better understand those quantities and their corresponding assumptions, let's imagine a following study. In this study we want to determine the effect of an experimental knee surgery on the activity level among patients with certain health condition. We have access to the data from a clinic in which surgeons used national guidelines to determine whether they will perform a surgery or not. According to those guidelines, patients who are at least $50$ years old are eligible for this surgery. This give an opportunity to employ RD approach. However, by a communication error the doctors at one ward were informed that patients who are at least $55$ years old are eligible. To fix the attention, we focus on the interpretation of the local average treatment effect at $55$. Let us consider the following scenarios. In the first scenario, there are no differences between wards both in terms of surgeons skills and the profile of patients they receive. Then, the average treatment effect we estimate is given by quantity \eqref{res3_app}, and it is identified for the population of all compliers aged 55 who were admitted to the clinic.  In the second scenario, the ward that received incorrect information has a higher proportion of junior surgeons compared to other wards. However, the profile of the patients does not differ between the wards due to standardized admission procedures. Then the local average treatment effect is given by the quantity \eqref{res2_app}, and it corresponds to the population of compliers aged 55 who were treated by, on average, less experienced surgeons. In the third scenario, because the doctors on the misinformed ward are less experienced, they tend to be assigned slightly easier cases compared to those on other wards. Then the local average treatment effect is given by \eqref{res1_app}, and it corresponds to the average treatment effect for compliers aged 55 who were in better condition than the average complier and were treated by, on average, less experienced surgeons.
\section{Algorithm implementation}
We program our models in JAGS, which makes it easy for other users to modify. We approximate the discontinuities at $\fk{l}$ and $\fk{r}$ in the outcome function with a sigmoid function.
\begin{align*}
    \Tilde{f}(x)=\begin{cases}
        {\fl{0}}+{\fl{1}}(x-c)+g(100(\fk{l}-x))(\fl{2}(x-c)^2+\fl{3}(x-c)^3), & \text{ for } x<c,\\
        {\fr{0}}+{\fr{1}}(x-c)+g(100(x-\fk{r}))(\fr{2}(x-c)^2+\fr{3}(x-c)^3), & \text{ for } c\leq x,
        \end{cases}
\end{align*}
where $g(x)$ is inverse logit function. It is important to notice that the order in which variables are introduced in the JAGS code may influence the convergence speed, and therefore the overall quality of the posterior samples. We initiate the MCMC chains based on the given data. In particular, in the preliminary step we fit two constant functions to the treatment data to get a rough estimate of the cutoff location. Then, randomly selected posterior samples are used as initial values of the cutoff $c$.

In terms of computational time, we obtained the following results (Apple M2 Pro, 16 GB RAM, macOS Sonoma 14.2.1). ART full dataset (6819 data points, 4 parallel chains, burnin: 40000, samples: 25000): 13 h 11 min; ART trimmed dataset (3507 data points, 4 parallel chains, burnin: 40000, samples: 25000): 5 h 30 min;  chemotherapy dataset (1923 data points, 4 parallel chains, burnin: 30000, sample: 25000): 39 min; single simulation for fuzzy design with unknown cutoff (500 data points, 4 parallel chains, burnin: 10000, adapt: 1000, sample: 25000): 2.5 min; single simulation for sharp design (500 data points, 4 parallel chains, burnin: 10000, adapt: 1000, sample: 25000): 45 sec.

\section{Parameter constraints for LoTTA treatment probability function}\label{App:ParamConst}

To ensure that the function $p$ is increasing, takes values between $0$ and $1$ and has discontinuity of size $j$ at $c$, where $j\geq \eta$, we solve the following  system of inequalities and equations.
\begin{equation*}
\begin{aligned}[c]
&\pa{i}{l}\geq 0 \text{ for } i=1,2,  \\
&\pa{2}{L}I_1+\pb{2}{L}\geq 0, \\
&\pa{2}{L}\pk{l}+\pb{2}{L}= \pa{1}{L}\pk{l}+\pb{1}{L},\\
&\pa{1}{L}c+\pb{1}{L} \leq 1-j,
\end{aligned}
\quad
\begin{aligned}[c]
&\pa{i}{r}\geq 0 \text{ for } i=1,2, \\
&\pa{1}{r}c+\pb{1}{r}= \pa{1}{l}c+\pb{1}{l}+j,\\
&\pa{1}{r}\pk{r}+\pb{1}{r}=\pa{2}{r}\pk{r}+\pb{2}{r},\\
&\pa{2}{r}\pk{r}+\pb{2}{r}\leq 1.
\end{aligned}
\\
\end{equation*}

\end{document}